\documentclass[conference]{IEEEtran}
\IEEEoverridecommandlockouts
\pdfoutput=1

\usepackage[utf8]{inputenc}
\usepackage{graphicx}
\usepackage{amsmath, amssymb}
\usepackage{booktabs}
\usepackage{hyperref}
\usepackage{cite}
\usepackage{enumitem}
\usepackage{multirow}
\usepackage{tabularx}
\usepackage{url}
\usepackage{svg}
\usepackage{xcolor}

\setlist[itemize]{noitemsep, topsep=0pt}
\setlist[enumerate]{noitemsep, topsep=0pt}

\title{Systematization of Knowledge: Security and Safety in the Model Context Protocol Ecosystem}
\author{
    \IEEEauthorblockN{Shiva Gaire\IEEEauthorrefmark{1}, Srijan Gyawali\IEEEauthorrefmark{1}, Saroj Mishra\IEEEauthorrefmark{1}, Suman Niroula\IEEEauthorrefmark{1}, Dilip Thakur\IEEEauthorrefmark{1}, and Umesh Yadav\IEEEauthorrefmark{1}}
    \IEEEauthorblockA{\textit{Tribhuvan University}: mail@shivagaire.com.np, gyawalisrijan01@gmail.com}
    \IEEEauthorblockA{\textit{University of North Dakota}: saroj.mishra773@gmail.com}
    \IEEEauthorblockA{\textit{Youngstown State University}: sum.nir1@gmail.com}
    \IEEEauthorblockA{\textit{University of Missouri}: dileepthakur87@gmail.com}
    \IEEEauthorblockA{\textit{University of Toledo}: yadav.umesh0518@gmail.com}
}

\begin{document}

\maketitle

\begin{abstract}
\begingroup
\renewcommand\thefootnote{}
\footnotetext{\IEEEauthorrefmark{1}All authors contributed equally to this work.}
\endgroup
The Model Context Protocol (MCP) has emerged as the de facto standard for connecting Large Language Models (LLMs) to external data and tools, effectively functioning as the "USB-C for Agentic AI." While this decoupling of context and execution solves critical interoperability challenges, it introduces a profound new threat landscape, where the boundary between epistemic errors (hallucinations) and security breaches (unauthorized actions) dissolves. This Systematization of Knowledge (SoK) aims to provide a comprehensive taxonomy of risks in the MCP ecosystem, distinguishing between adversarial security threats (e.g., indirect prompt injection, tool poisoning) and epistemic safety hazards (e.g., alignment failures in distributed tool delegation). We analyze the structural vulnerabilities of MCP primitives, specifically Resources, Prompts, and Tools, and demonstrate how "context" can be weaponized to trigger unauthorized operations in multi-agent environments. Furthermore, we survey state-of-the-art defenses, ranging from cryptographic provenance (ETDI) to runtime intent verification, and conclude with a roadmap for securing the transition from conversational chatbots to autonomous agentic operating systems.
\end{abstract}


\section{Introduction}

\subsection{Background and Motivation}
The field of Artificial Intelligence is undergoing a paradigmatic shift from \textit{Conversational AI}—where models generate text in isolation—to \textit{Agentic AI}, where models perceive, reason, and act upon the external world. This transition requires a standardized connective tissue to link probabilistic Large Language Models (LLMs) with deterministic digital systems. The \textbf{Model Context Protocol (MCP)}, introduced in late 2024, has emerged as this standard, effectively serving as the ``USB-C for AI applications'' by abstracting the complexities of data retrieval and tool execution into a unified open protocol \cite{mcp_spec, cross2025smcp}.

The adoption of MCP solves a critical interoperability bottleneck, famously known as the ``$M \times N$ integration problem,'' allowing any model to connect to any data source without bespoke adapters \cite{hou2025mcplandscape}. However, this architectural decoupling introduces profound security implications. By standardizing the interface between an LLM and local files, databases, and remote APIs, MCP significantly expands the attack surface of AI systems. It transforms the LLM from a passive text processor into an active system component with shell-level privileges, capable of executing actions based on potentially untrusted context.

As MCP adoption accelerates in enterprise environments—powering IDEs, data pipelines, and customer support agents—the industry faces a critical knowledge gap. While individual vulnerabilities like prompt injection are well-documented, there is no comprehensive framework understanding how these threats manifest in a decentralized, protocol-driven ecosystem where control flow is determined by semantic context rather than code.

\subsection{Problem Statement: Security vs. Safety in MCP}
The core challenge in securing MCP ecosystems lies in the convergence of \textit{security} and \textit{safety} failures. In traditional software, these domains are distinct: security protects against malicious adversaries (e.g., SQL injection), while safety protects against unintended system behaviors (e.g., race conditions). In MCP, this distinction blurs.

A ``security'' breach, such as an attacker injecting a malicious document into a company's knowledge base (Indirect Prompt Injection), can trigger a ``safety'' failure, where the model honestly but mistakenly believes it is authorized to delete a database. Conversely, a safety failure, such as model hallucination regarding a tool's parameters, can lead to a security breach where sensitive data is exfiltrated to a public log \cite{radosevich2025mcpsafetyaudit}.

Current defense mechanisms are ill-equipped for this duality. Traditional firewalls cannot inspect the semantic intent of a JSON-RPC message, and LLM safety filters cannot see the downstream consequences of a tool execution. This paper argues that securing MCP requires a unified threat model that treats context availability and execution privilege as inextricably linked variables.

\subsection{Scope of the Survey}
This Systematization of Knowledge (SoK) focuses on the unique risks introduced by the \textbf{Model Context Protocol ecosystem}. Our analysis encompasses:
\begin{itemize}
    \item \textbf{Protocol Primitives:} Vulnerabilities inherent in the design of Resources, Prompts, and Tools as defined in the MCP specification \cite{mcp_spec}.
    \item \textbf{Topology Risks:} Threats arising from the distributed nature of Host-Client-Server interactions, including supply chain risks in open tool registries.
    \item \textbf{Intersection of Threats:} We specifically exclude general LLM adversarial attacks (e.g., weight poisoning) unless they directly impact the protocol's integrity or execution flow.
\end{itemize}

\subsection{Contributions of this Paper}
To our knowledge, this is the first academic survey to systematize the risks of the Model Context Protocol. Our contributions are as follows:

\begin{enumerate}
    \item \textbf{Unified Vulnerability Taxonomy:} We propose a novel taxonomy (Table \ref{tab:mcp-vulnerability-taxonomy}) that distinguishes between \textit{Adversarial Security Threats} (e.g., tool masquerading, context poisoning) and \textit{Epistemic Safety Hazards} (e.g., alignment failures in tool delegation).
    \item \textbf{Structural Analysis of MCP Primitives:} We analyze how the decoupling of "Context" (Resources) and "Action" (Tools) creates new classes of vulnerabilities, such as Cross-Primitive Escalation, where read-only access is weaponized to trigger write-actions \cite{guo2025systematic}.
    \item \textbf{Survey of Emerging Defenses:} We synthesize state-of-the-art mitigation strategies, moving beyond basic prompt engineering to architectural solutions like the Enhanced Tool Definition Interface (ETDI) \cite{bhatt2025etdi} and kernel-level session isolation \cite{bui2020gvisor}.
    \item \textbf{Forensic Case Studies:} We reconstruct real-world incidents, such as the Supabase data leak \cite{pomerium2025airoot}, to derive actionable lessons for enterprise deployment.
\end{enumerate}

\subsection{Organization of the Paper}
The remainder of this paper is organized as follows: Section \ref{sec:overview} provides a technical overview of the MCP architecture. Section \ref{sec:threat-landscape} defines the threat landscape and adversarial actors. Sections \ref{sec:security-challenges} and \ref{sec:safety-challenges} detail the specific security and safety challenges, respectively. Section VI surveys mitigation strategies and architectural defenses. Section VII outlines open research directions, and Section VIII presents case studies of recent MCP-related incidents. Finally, Section IX concludes with a roadmap for secure adoption.
\section{Overview of the Model Context Protocol (MCP)}
\label{sec:overview}

The Model Context Protocol (MCP) establishes a standardized open protocol that decouples AI models from their data sources and tools. By abstracting these connections into a client-host-server topology, MCP aims to solve the interoperability challenges inherent in connecting Large Language Models (LLMs) to local and remote ecosystems \cite{mcp_spec}.

\subsection{Evolution of Context Protocols in AI Systems}

The integration of external context into AI systems has evolved through three distinct phases. Initially, developers relied on \textit{bespoke glue code} and static context injection, where retrieval logic was hard-coded into the application layer. This unscalable approach led to the ``$M \times N$ integration problem,'' where every model provider ($M$) required custom connectors for every data source ($N$) \cite{hou2025mcplandscape}.

The second phase introduced \textit{proprietary plugin ecosystems} (e.g., OpenAI Plugins), which standardized tool definitions but locked developers into specific model vendors. MCP represents the third phase: a \textit{universal open standard} that operates over local and remote transports (Stdio, SSE), allowing any model to connect to any server without vendor-specific adapters \cite{mcp_architecture_blog}.

\subsection{MCP Architecture and Design Principles}

\begin{figure}[htbp]
    \centering
    \includegraphics[width=\linewidth]{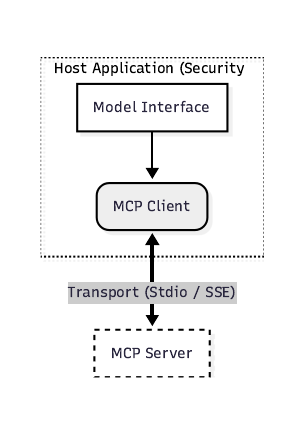}
    \caption{MCP Architecture. The host application acts as the security boundary, mediating interactions between the Model Interface and the external MCP Server \cite{mcp_spec}.}
    \label{fig:mcp_arch}
\end{figure}

The architecture is founded on a \textbf{Client-Host-Server} model designed to run locally or remotely. The design prioritizes security boundaries by ensuring the LLM never connects directly to a data source; instead, all interactions are mediated by the host application \cite{mcp_spec}.

Key design principles include:
\begin{itemize}
    \item \textbf{Transport Agnosticism:} The protocol runs over standard input/output (Stdio) for local process isolation or Server-Sent Events (SSE) for remote connections.
    \item \textbf{Capability Negotiation:} Connections begin with a handshake where Client and Server declare their supported features (e.g., resources, logging, prompts) before exchanging data.
    \item \textbf{JSON-RPC 2.0:} The message format relies on a stateless, lightweight remote procedure call standard, ensuring compatibility across programming languages \cite{mcp_spec}.
\end{itemize}

\subsection{Core Components and Data Flow}

\begin{figure}[htbp]
    \centering
    \includegraphics[width=\linewidth]{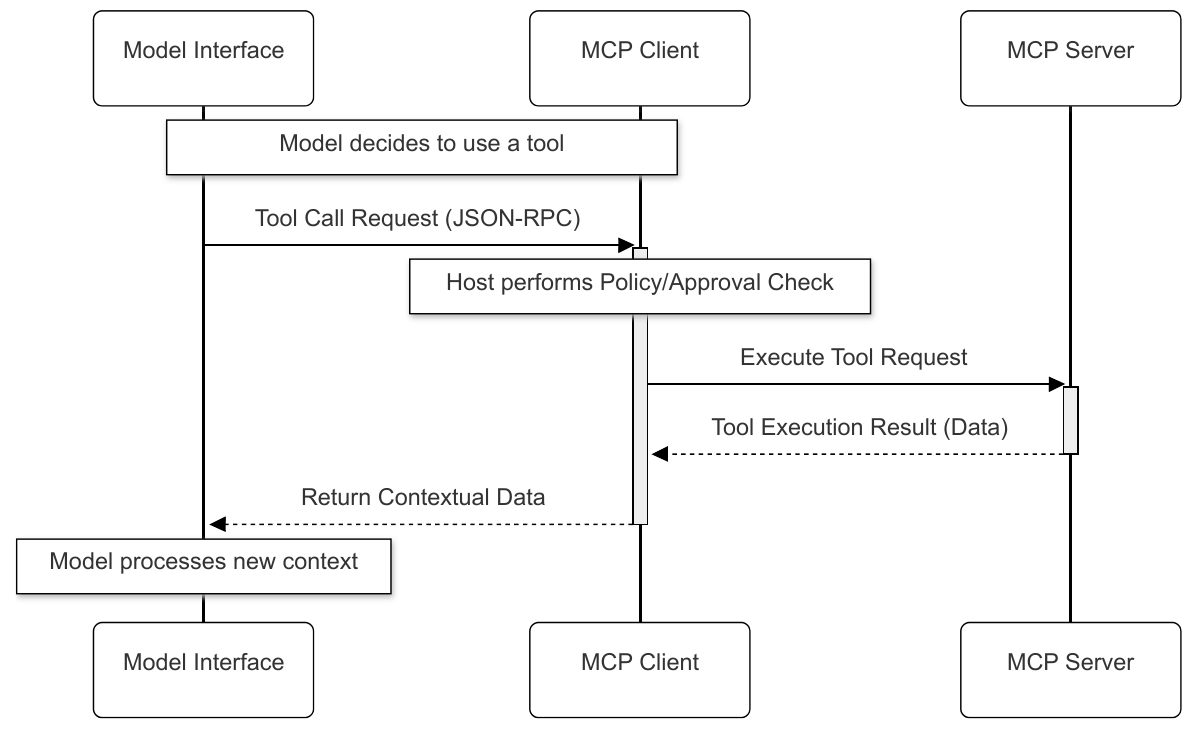}
    \caption{MCP Data Flow. The sequence demonstrates how a Model Interface request is routed through the Client, approved by the Host Policy check, and executed by the Context Source.}
    \label{fig:mcp_dataflow}
\end{figure}

\subsubsection{Model Interfaces}
The ``Model Interface'' in MCP is the abstraction layer managed by the host application (e.g., Claude Desktop, IDEs). It is responsible for the sampling loop: sending the conversation history to the LLM, parsing the LLM's output for tool call requests, and serializing those requests into MCP-compliant JSON-RPC messages. Crucially, this interface creates a buffer between the probabilistic nature of the model and the deterministic nature of the protocol, ensuring that model hallucinations do not directly corrupt the protocol state \cite{mcp_architecture_blog}.

\subsubsection{Context Sources and Connectors}
In the MCP ecosystem, context sources are encapsulated as ``MCP Servers.'' These servers act as connectors that expose data via two primary primitives:
\begin{itemize}
    \item \textbf{Resources:} Passive data streams (identified by URIs like \texttt{file:///logs/error.txt}) that the model can read. These function similarly to GET requests in REST but include subscription capabilities for real-time updates.
    \item \textbf{Prompts:} Server-defined templates that pre-package resources and instructions. These allow connectors to define ``best practice'' workflows (e.g., a ``Git Commit'' prompt that automatically grabs the diff and asks for a message) \cite{mcp_spec}.
\end{itemize}

\subsubsection{Clients and Applications}
The ``MCP Client'' is the protocol implementation embedded within the host application. While the Host manages the user interface and process lifecycle, the Client handles the strict protocol mechanics: maintaining the connection, routing messages, and handling error states. The Client acts as an aggregator, capable of connecting to multiple Servers simultaneously and presenting a unified list of tools to the Application layer. This aggregation enables the ``bring your own tools'' paradigm central to MCP \cite{hou2025mcplandscape}.

\subsubsection{Governance and Policy Layers}
Governance is enforced at the protocol edge via the capabilities handshake and runtime permissions. Unlike web APIs where authentication is often handled via headers, MCP relies on \textit{process-based access control}. The host application serves as the policy enforcement point, intercepting tool execution requests (e.g., ``Delete File'') and requiring user or policy approval before passing the command to the Server. Current research suggests this Human-in-the-Loop (HITL) mechanism is the primary (and often single) line of defense against agentic risks \cite{radosevich2025mcpsafetyaudit, bhatt2025etdi}.

\subsection{Comparison with Traditional Protocol Ecosystems}

MCP differs from traditional middleware like REST or gRPC by embedding semantic intent into the connection.

\begin{figure}[htbp]
    \centering
    \includegraphics[width=\linewidth]{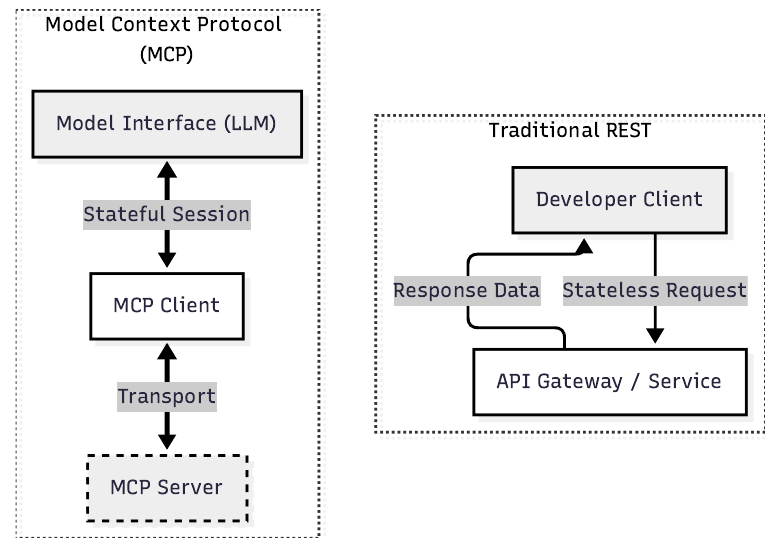}
    \caption{Comparison of Control Flow. Traditional REST APIs (left) rely on stateless, developer-driven requests. MCP (right) relies on stateful, intent-driven sessions where the model determines execution paths.}
    \label{fig:mcp_vs_rest}
\end{figure}

\begin{table}[h]
\centering
\caption{Comparison of MCP vs. Traditional Protocols}
\label{tab:protocol-comparison}
\begin{tabularx}{\linewidth}{l X X}
\toprule
\textbf{Feature} & \textbf{REST / OpenAPI} & \textbf{Model Context Protocol} \\
\midrule
\textbf{Topology} & Stateless Request/Response & Stateful Session (Stdio/SSE) \\
\textbf{Discovery} & Static Schema (Swagger) & Dynamic Capability Negotiation \\
\textbf{Control Flow} & Client-Driven & Model-Intent Driven \\
\textbf{Security} & Endpoint Authentication & Host-mediated Process Isolation \\
\bottomrule
\end{tabularx}
\end{table}

While REST APIs are designed for deterministic developer interactions, MCP is optimized for non-deterministic model interactions, where the ``caller'' (the LLM) determines the execution path based on context \cite{hou2025mcplandscape}.

\subsection{Role of MCP in AI Integration and Multi-Agent Systems}

MCP serves as the interoperability layer for Agentic AI. By standardizing the tool interface, it solves the fragmentation problem in Multi-Agent Systems (MAS). In an MCP-enabled MAS, agents can query the capabilities of other agents (exposed as Servers) and hand off tasks dynamically. For example, a ``Coder Agent'' can connect to a ``Database Agent'' via MCP to inspect a schema, treating the Database Agent's tools as its own context. This composability is essential for scaling from single-task bots to complex, multi-modal agent ecosystems \cite{datta2025agentic}.
Recent comparative studies position MCP as the foundational tier for tool execution, distinct from but complementary to high-level coordination standards like the Agent-to-Agent (A2A) and Agent Network Protocols (ANP) \cite{ehtesham2025survey}.

\section{Threat Landscape in the MCP Ecosystem}
\label{sec:threat-landscape}

The Model Context Protocol (MCP) represents a new paradigm for connecting large language models (LLMs) with external tools and contextual resources. While this integration expands capability and flexibility, it simultaneously broadens the attack surface. The decentralized structure - comprising independent Hosts, Clients, and Servers - means that vulnerabilities in one component can propagate across others, converting isolated weaknesses into systemic risks \cite{guo2025systematic,piazza2025mcpnightmare,hou2025mcplandscape}.

\subsection{Adversarial Actors and Attack Vectors}

The MCP environment attracts a wide spectrum of adversaries. Malicious developers may publish deceptive MCP servers or modify tool definitions to exfiltrate data after installation, a pattern analyzed as tool squatting and ``rug-pull'' behavior \cite{bhatt2025etdi}. Supply-chain attackers exploit the lack of centralized distribution by introducing tampered installers or rogue updates that silently alter tool functions or metadata \cite{hou2025mcplandscape,bhatt2025etdi,shapira2025mcpsecurity}. Insider threats within organizations can also misuse legitimate MCP agents or servers to reach confidential enterprise data, for example by over-scoped tools and misconfigured Resource access \cite{florencio2025mcpredhat,shapira2025mcpsecurity}. 

Beyond individual insiders, sophisticated adversaries can take advantage of the same structural weaknesses. Existing MCP security work documents how poorly authenticated or spoofed servers, over-privileged tools, and weak integrity checks enable remote code execution, token theft, and lateral movement across systems \cite{guo2025systematic,radosevich2025mcpsafetyaudit,shapira2025mcpsecurity}. In such an environment, state-sponsored or organized actors could, in principle, embed these techniques into large-scale disinformation or espionage workflows that operate through MCP-connected agents, even though current papers emphasize the technical exploit chains rather than specific geopolitical campaigns.

These adversaries take advantage of the protocol's limited integrity verification, mutable server-side logic, and the absence of uniform, continuous validation mechanisms across implementations \cite{guo2025systematic,florencio2025mcpredhat,shapira2025mcpsecurity}. Because MCP grants every server autonomy over tool exposure and scope, a single compromised endpoint can cascade through dependent agents and services, demonstrating the fragility of distributed trust in early MCP ecosystems \cite{guo2025systematic,hou2025mcplandscape,shapira2025mcpsecurity}.

\subsection{Taxonomy of Threats: Security vs.\ Safety}

MCP threats can be organized along two complementary axes: \emph{security} (unauthorized access or modification) and \emph{safety} (unintended but harmful outcomes).

\begin{table*}[t]
\centering
\resizebox{\linewidth}{!}{

\label{tab:security-safety-taxonomy}
\begin{tabular}{p{3cm} p{5.2cm} p{3cm} p{3.5cm}}
\toprule
\textbf{Dimension} & \textbf{Security focus} & \textbf{Safety focus} \\
\midrule
Integrity 
& Tool or Resource tampering, unauthorized context injection, tool poisoning, and abuse of MCP message flows \cite{guo2025systematic,piazza2025mcpnightmare,shapira2025mcpsecurity}. 
& Model misinterpretation or grounding failures that cause tools to be invoked in harmful ways or with incorrect parameters (see Section~\ref{sec:safety-challenges}). \\[0.4em]

Confidentiality 
& Data exfiltration via malicious or spoofed MCP servers, exposure of tokens or session identifiers, and cross-context leaks between tools and workflows \cite{guo2025systematic,shapira2025mcpsecurity}. 
& Accidental leakage of private information during reasoning, for example when over-broad tool scopes surface sensitive data into the model context or logs \cite{florencio2025mcpredhat,shapira2025mcpsecurity}. \\[0.4em]

Availability 
& Denial-of-service and resource-exhaustion conditions targeting MCP servers and connected LLM services, including unbounded consumption patterns that degrade or disable tool workflows \cite{narajala2025enterprise,sauter2024unbounded}.
& Over-automation or runaway execution loops that consume resources, reduce human oversight, or trigger repeated tool calls without meaningful user control \cite{sauter2024unbounded}. \\
\bottomrule
\end{tabular}}
\caption{Security and safety dimensions in MCP deployments.}
\end{table*}

Security incidents generally involve unauthorized manipulation (for example, tampered tool definitions, command injection, or stolen credentials), whereas safety failures arise when MCP systems follow syntactically valid instructions that nonetheless lead to undesirable consequences. In practice, these dimensions intersect: compromised security often precipitates safety breakdowns, and conversely, weak safety constraints can make security exploits easier to weaponize.

\subsection{Unique Threats in Context-Driven Protocols}

Unlike traditional APIs such as REST or gRPC, MCP combines model reasoning with executable control, creating a semantic layer that is vulnerable to meaning-based manipulation. Several attack classes rely on this context-driven design \cite{guo2025systematic,hou2025mcplandscape,florencio2025mcpredhat,shapira2025mcpsecurity}:

\begin{itemize}
  \item \textbf{Indirect tool injection and prompt injection.} Malicious instructions are embedded in contextual data (for example, documents, emails, or tool outputs) and then forwarded into tools or servers via the MCP client, allowing attackers to manipulate agent behavior or tool invocation paths \cite{guo2025systematic,shapira2025mcpsecurity}.
  \item \textbf{Context poisoning and cross-context contamination.} Malicious artifacts persist across tools, sessions, or agents, for instance through shared file-based context or reused tool registries, enabling chain attacks that exploit shared context and confuse trust boundaries \cite{guo2025systematic,shapira2025mcpsecurity}.
  \item \textbf{Model-switch and server shadowing.} Attackers register MCP servers or tools that mimic trusted ones, or exploit weak naming/namespace controls, to redirect tool calls to malicious implementations (sometimes described as spoofed or ``shadow'' MCP servers) \cite{hou2025mcplandscape,shapira2025mcpsecurity}.
  \item \textbf{Protocol abuse via crafted MCP messages.} Manipulated MCP requests and responses can encode unintended commands or parameters that agents treat as legitimate tool invocations, especially when input validation and policy checks are weak \cite{guo2025systematic,radosevich2025mcpsafetyaudit,shapira2025mcpsecurity}.
\end{itemize}

Because MCP standardizes tool and Resource interaction, a single successful exploit pattern, such as a prompt-injection template or a malicious server configuration, can be replicated across many implementations. This shared surface makes defensive containment particularly challenging in multi-tenant and multi-server ecosystems \cite{guo2025systematic,shapira2025mcpsecurity}.

\subsection{Lessons from Related Protocol Security Incidents}

Historical precedents illustrate how open ecosystems tend to evolve from rapid innovation toward increasingly strict trust and verification controls. Browser extensions and package registries such as npm and PyPI initially thrived on openness, but supply-chain abuse and malicious packages eventually forced providers to adopt stronger signing, publisher verification, and reputation mechanisms. Recent MCP security analyses explicitly connect these lessons to MCP package ecosystems, highlighting risks such as tool name collisions, malicious MCP servers distributed via open registries, and weak integrity checks on MCP packages \cite{shapira2025mcpsecurity}.

Analyses by both academic and industry groups emphasize that future MCP deployments will need practices aligned with zero-trust approaches \cite{hou2025mcplandscape,florencio2025mcpredhat,narajala2025enterprise,shapira2025mcpsecurity}. Recommended controls include mandatory and identity-bound authentication between Hosts, Clients, and Servers; tightly scoped authorization using OAuth or equivalent mechanisms; integrity protections such as signed MCP packages or tool definitions where feasible; and provenance or audit trails that make each context element and tool invocation verifiable before execution. Without such standards, MCP's modular design risks allowing threats to propagate faster than defenses can adapt.

Ultimately, the MCP threat landscape shows that the same properties that enable powerful AI coordination - interoperability, extensibility, and automation - also introduce unprecedented risk. Addressing these issues requires cross-disciplinary governance that bridges AI safety research with classical cybersecurity engineering, and treats MCP security as an evolving, ecosystem-level problem rather than a purely local implementation concern.

\section{Security Challenges in the Model Context Protocol (MCP)}
\label{sec:security-challenges}
\begin{figure}[htbp]
    \centering
    \includegraphics[width=\linewidth]{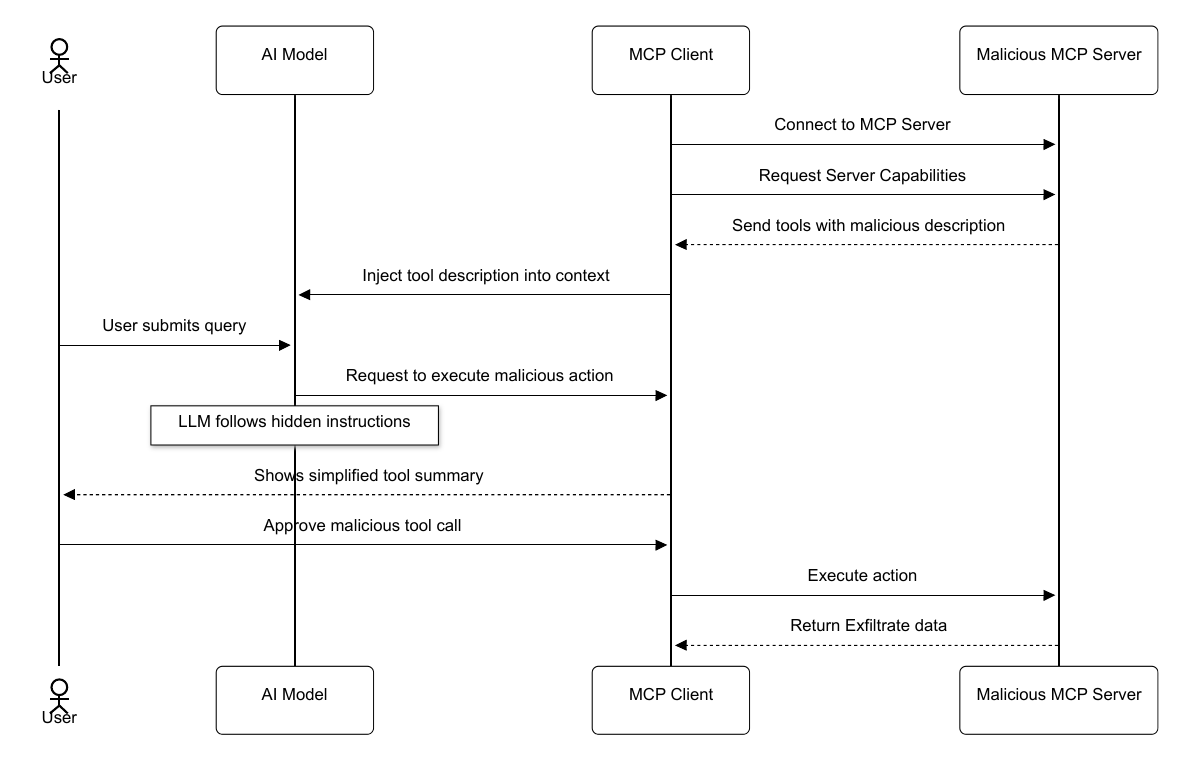}
    \caption{Context Poisoning Attack Flow. An attacker injects a malicious tool description into the context, causing the LLM to unknowingly execute unauthorized actions despite user approval checks.}
    \label{fig:context_poison}
\end{figure}

The Model Context Protocol (MCP) is still an emerging standard whose security posture remains underdeveloped. Unlike mature middleware frameworks such as REST or gRPC, which separate data transport, authentication, and execution, MCP merges reasoning and control flows within a shared semantic context \cite{guo2025systematic,hou2025mcplandscape}. This enables fluid coordination between LLMs and external tools but simultaneously blurs traditional trust boundaries. In many MCP implementations, context, metadata, and executable instructions coexist without strong isolation, allowing malicious actors to exploit semantic ambiguity or tool overreach. The lack of uniform guidelines or centralized accountability further amplifies these risks: many vendors delegate hardening to integrators or end users, rather than enforcing baseline safeguards \cite{florencio2025mcpredhat,piazza2025mcpnightmare}. Traub notes that ambiguous terminology—particularly the conflation of local executables with remote MCP servers—creates dangerous misconceptions about privilege boundaries and exposure surfaces \cite{traub2025terminology}.

These vulnerabilities are deeply interrelated rather than isolated. Context poisoning can enable unauthorized context injection and data leakage; weak permission boundaries allow tools to read sensitive files; insecure configuration or update paths can lead to supply-chain compromise; and malicious server impersonation can redirect tool calls. Because MCP couples model reasoning with tool execution, the protocol inherits both software vulnerabilities and linguistic/contextual ones. Unlike conventional APIs, where attacks primarily exploit code-level flaws, MCP introduces an attack surface rooted in interpretation: what the model understands can be as dangerous as what the code executes.

\subsection{MCP Security Vulnerability Taxonomy}

Table~\ref{tab:mcp-vulnerability-taxonomy} provides a detailed classification of the security and safety risks within the MCP ecosystem, categorized by impact and execution phase.

\begin{table*}[t]
\centering
\caption{MCP Security Vulnerability Taxonomy}
\label{tab:mcp-vulnerability-taxonomy} 

\resizebox{\linewidth}{!}{
    \begin{tabular}{p{3cm} p{5.2cm} p{3cm} p{3.5cm}}
    \toprule
    \textbf{Category} & \textbf{Impact} & \textbf{Phase} & \textbf{Evidence Source} \\
    \midrule
    Context Poisoning &
    Data leakage; unauthorized command execution; confidentiality breach &
    Execution &
    Guo et al.\ (2025); Beurer\textendash Kellner \& Fischer (2025) \\[0.4em]

    Prompt Injection &
    Unauthorized actions; model manipulation &
    Execution &
    Florencio (2025); Guo et al.\ (2025) \\[0.4em]

    Unauthorized Context Injection &
    Stealthy tool override; integrity breach &
    Install/Update/Exec &
    Guo et al.\ (2025) \\[0.4em]

    Data Leakage \& Privacy Risks &
    Credential leakage; session exposure; file exfiltration &
    Execution &
    Piazza (2025); Guo et al.\ (2025); Beurer\textendash Kellner \& Fischer (2025) \\[0.4em]

    Cross-Session Contamination &
    Infectious attacks; multi-user integrity breach &
    Execution &
    Guo et al.\ (2025); Radosevich \& Halloran (2025) \\[0.4em]

    Supply-Chain \& Model-Switch Attacks &
    Rug-pull attacks; server impersonation; unauthorized tool mutation &
    Install/Update &
    Hou et al.\ (2025); Bhatt et al.\ (2025) \\[0.4em]

    Protocol Abuse \& Name Collisions &
    Command overlap; confused-deputy access &
    Execution &
    Piazza (2025); Hou et al.\ (2025); Florencio (2025); Shapira (2025) \\[0.4em]

    DoS \& Resource Exhaustion &
    System unresponsiveness; cost escalation &
    Execution &
    Narajala \& Habler (2025); Sauter (2024); Guo et al.\ (2025) \\
    \bottomrule
    \end{tabular}
}
\end{table*}

\subsection{Context Poisoning and Prompt Injection}

Context poisoning embeds malicious instructions in tool metadata or schema so the LLM executes unintended actions. Guo et al.\ describe these as \emph{tool poisoning attacks}, where malicious arguments or hidden behaviors in the tool description bypass user awareness \cite{guo2025systematic}. Beurer-Kellner and Fischer demonstrate that agents often ``blindly rely'' on docstrings: hidden steps inside metadata can cause file access or exfiltration through an otherwise benign-looking tool \cite{beurerkellner2025toolpoisoning}. Because many MCP clients do not display full tool definitions, the model may unknowingly execute harmful operations.

Prompt injection instead manipulates user-facing context (documents, pasted text, retrieved content). MCP treats contextual inputs as authoritative; if an attacker embeds obfuscated instructions in upstream content, the agent may redirect tool calls or modify behaviour \cite{florencio2025mcpredhat}. Both vectors exploit the fact that MCP tightly couples contextual understanding with execution.

\subsection{Unauthorized Context Injection}

Unauthorized context injection occurs when malicious data, commands, or tool definitions are introduced into the MCP environment during installation, updates, or runtime. Guo et al.\ categorize this as \emph{indirect tool injection}: an attacker spoofs an installer or update channel to insert hidden tool entries or override existing ones \cite{guo2025systematic}. Because installation and configuration can be complex, third-party installers without strong controls may introduce manipulated payloads. Malicious servers may also shadow trusted ones, providing altered tools under familiar names. Without explicit server authentication or integrity verification, these injections can grant attackers silent control.

\subsection{Data Leakage and Privacy Risks}

MCP servers often expose tools with access to local files, credentials, or logs. If a tool or server is compromised, it may exfiltrate such data. Piazza highlights that some MCP servers expose session identifiers or sensitive metadata without strict controls, creating opportunities for leakage or session hijacking \cite{piazza2025mcpnightmare}. Guo et al.\ show that file-reading behaviors embedded in tool metadata can cause unintended disclosure \cite{guo2025systematic}. Beurer-Kellner \& Fischer further demonstrate that poisoned docstrings can lead an agent to read configuration files or sensitive user data as part of tool invocation \cite{beurerkellner2025toolpoisoning}. Without sandboxing or least-privilege permissions, confidentiality risks are substantial.

\subsection{Cross-Session Contamination}

Operationally, MCP agents may maintain shared state across sessions. Guo et al.\ demonstrate that poisoned metadata or artifacts can propagate across tools and tasks, producing ``infectious'' effects \cite{guo2025systematic}. Radosevich \& Halloran describe scenarios where corrupted public data retrieved through MCP tools leads to execution of attacker-provided instructions in later sessions \cite{radosevich2025mcpsafetyaudit}. Without strict session isolation, contamination in one user’s workflow may compromise others.

\subsection{Supply-Chain and Model-Switch Attacks}
\begin{figure}[htbp]
    \centering
    \includegraphics[width=\linewidth]{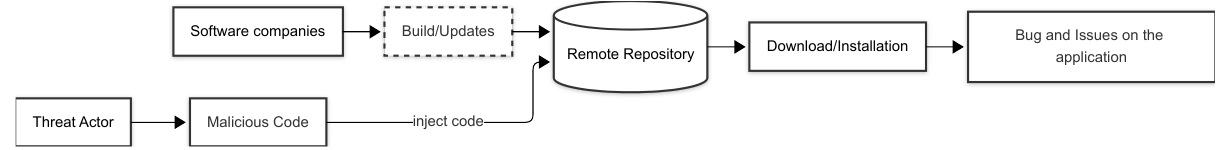}
    \caption{Supply Chain Attack Vector in MCP. Threat actors inject malicious code into remote repositories, which are then downloaded by unwitting users during build or update cycles, compromising the application environment.}
    \label{fig:supply_chain_attack}
\end{figure}

Because MCP depends on distributed servers and community tooling, it is highly susceptible to supply-chain compromise. Hou et al.\ note risks of spoofed or malicious distributions (e.g.\ manipulated installers or replacement packages) \cite{hou2025mcplandscape}. Bhatt et al.\ identify four structural causes behind \emph{rug-pull} attacks: mutable server-side logic, lack of continuous integrity checks, absence of re-approval triggers, and exploitation of established trust \cite{bhatt2025etdi}. Model-switching attacks occur when attackers register servers mimicking trusted ones to redirect tool calls.

\subsection{Insecure Serialization and Protocol Abuse}

MCP servers often rely on structured messages (e.g., JSON transport). Piazza highlights that many servers lack integrity checks, allowing manipulated messages or unauthorized modifications to flow through unverified \cite{piazza2025mcpnightmare}. Shapira documents attacks leveraging namespace collisions or command overlap in registries without isolation \cite{shapira2025mcpsecurity}. Hou et al.\ similarly warn that unchecked naming collisions could escalate in multi-tenant deployments \cite{hou2025mcplandscape}. These weaknesses permit protocol-level manipulation even when tool implementations are correct.

\subsection{Denial-of-Service and Resource Exhaustion}

DoS and resource-consumption attacks target MCP agents by forcing excessive tool calls, unbounded retrieval, or infinite loops. Narajala \& Habler classify such attacks as severe in enterprise settings, noting that recursive or repeated tool triggers can exhaust CPU and memory \cite{narajala2025enterprise}. Guo et al.\ observe that malicious payloads can significantly increase computational cost \cite{guo2025systematic}. Sauter describes \emph{unbounded consumption attacks} that cause runaway resource usage through crafted prompts rather than classical network flooding \cite{sauter2024unbounded}.

\subsection{Why MCP Security Is Vulnerable Today}

MCP adoption is rapid, yet governance and standardization lag behind. Many MCP clients do not expose full tool specifications, request permissions transparently, or enforce strict boundaries between context and execution. Third-party installers, shared contexts, lack of authentication, and unverified tool updates magnify exposure. Because MCP integrates both AI safety and classical cybersecurity concerns, securing it requires combined safeguards: authentication, sandboxing, provenance checks, tool validation, session isolation, continuous monitoring, and least-privilege enforcement. Without such measures, MCP deployments face heightened systemic risk.

\section{Safety Challenges in the Model Context Protocol Ecosystem}
\label{sec:safety-challenges}

While the security risks of MCP arise from adversarial compromise or unauthorized manipulation, safety risks emerge when the system produces correct actions according to the protocol but leads to unintended harmful outcomes. This distinction is widely emphasized in AI governance research, where safety failures stem from epistemic errors, misalignment between components, or inadequate oversight rather than explicit security breaches \cite{nist_airmf,eu_ai_act,owasp_llm_top10}.

The Model Context Protocol (MCP) introduces unique safety challenges due to its distributed and compositional architecture. Unlike monolithic LLM applications, MCP separates responsibilities across three primitives - Resources (read-only context), Tools (external actions), and Prompts (reusable templates) - each operating in distinct trust domains \cite{mcp_spec}. Because these primitives interact across independent Hosts, Clients, and Servers, the traditional ``single safety perimeter'' dissolves. As a result, failures in one primitive (e.g., low-fidelity contextual retrieval) can escalate into harmful actions executed by another.

These cascading interactions mirror patterns observed in retrieval-augmented generation (RAG) systems, where epistemic errors in upstream retrieval frequently propagate into downstream reasoning \cite{gao2023rag_survey}. In MCP, this propagation is amplified because the protocol can perform real-world actions, meaning that context errors or malicious instructions can transform into operational harm.

To analyze these risks systematically, this paper introduces a four-part taxonomy of safety challenges tailored to the MCP ecosystem (Table \ref{tab:mcp-safety-taxonomy}).

\begin{table}[t]
\centering
\caption{Mapping Safety Challenges to MCP Primitives and Risk Types}
\label{tab:mcp-safety-taxonomy}
\resizebox{\linewidth}{!}{
\begin{tabular}{p{2.5cm} p{2.7cm} p{2cm} p{4.5cm}}
\toprule
\textbf{Safety Category} & \textbf{Primary Primitive} & \textbf{Risk Type} & \textbf{Description} \\
\midrule
Epistemic Integrity & Resources & Epistemic & Grounding or hallucination due to fragmented or low-fidelity retrieval. \\
Adversarial Resilience & Resources \& Prompts & Adversarial & Hidden instructions or prompt injections embedded in external data. \\
Alignment Consistency & Host Model \& Tool Policies & Alignment & Policy conflicts or goal-pursuit misalignment leading to harmful delegation. \\
Systemic Governance & Human Oversight \& Regulation & Governance & HITL failures, accountability gaps, dual-use misuse. \\
\bottomrule
\end{tabular}}
\end{table}

\subsection{Hallucination and Grounding Failures}
Hallucination poses a significant epistemic safety risk in all LLM systems. In the MCP ecosystem, this risk is exacerbated by distributed data retrieval.

\subsubsection{Context Fragmentation and Distributed RAG Fidelity}
Conventional RAG systems suffer from ``context fragmentation,'' where splitting documents into small chunks breaks critical relationships \cite{liu2024lost_middle}. 

In MCP, Resources are retrieved from remote, independently managed Servers. This compounds fragmentation because fidelity loss occurs not only during chunking but also during transmission. If a remote Server employs a low-fidelity retrieval strategy, the Host receives isolated or incomplete context, making epistemic errors an inevitable outcome \cite{gao2023rag_survey}.

\subsubsection{Verification and Citation Challenges}
Reliable AI systems require traceability and the ability to cite sources \cite{ibm_hitl}. MCP complicates this because a Resource may represent aggregated content sourced from multiple Server-side documents \cite{mcp_spec}. Ensuring end-to-end traceability becomes a distributed systems challenge. Without standardized provenance metadata, citation verification becomes difficult in high-stakes environments.

\subsection{Adversarial Steering and Filter Evasion}
Adversarial steering manipulates an LLM through crafted inputs. In MCP, this typically manifests as Indirect Prompt Injection (IPI) \cite{owasp_llm_top10}.

\subsubsection{Indirect Prompt Injection}
IPI occurs when an LLM processes external inputs containing hidden instructions \cite{greshake2023more_than_you}. Because MCP Resources may contain arbitrary data such as files, Slack messages, or emails, these inputs can embed attacker-crafted directives intended to override system prompts.
\begin{figure}[htbp]
    \centering
    \includegraphics[width=\linewidth]{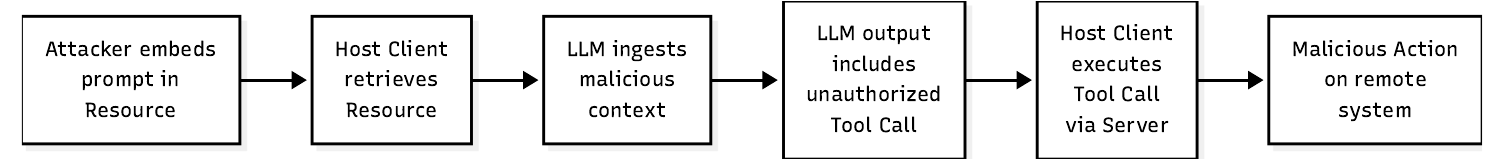}
    \caption{Indirect Prompt Injection Attack Flow. The attacker embeds a malicious prompt in a Resource (Red). When the Host Client retrieves this resource (Blue), the LLM ingests the malicious context (Orange) and unwittingly issues an unauthorized tool call (Purple), which is executed by the Host Client (Green), resulting in a malicious action on the remote system (Red).}
    \label{fig:prompt_injection_flow}
\end{figure}

\subsubsection{Cross-Primitive Execution Attacks}
The most severe IPI risk arises when an injected instruction escalates from epistemic harm to operational harm. Since MCP exposes Tools to LLMs, a malicious instruction within a Resource can trigger a destructive Tool action on a remote Server. 

This demonstrates a systemic safety gap in distributed environments where low-trust Resources can influence high-privilege Tool Servers.

\subsection{Policy Conflicts Across Components}
Alignment failures occur when the model’s objective function diverges from human intent. MCP complicates alignment because safety policies are distributed across independent components.

\subsubsection{Distributed Preference Alignment}
Host LLMs often rely on preference alignment methods such as RLHF. External Servers, however, operate under independent functional policies. When the Host delegates decisions to a Server, policy misalignment may allow the remote component to circumvent the Host’s guardrails. Leike et al. highlight that scaling alignment requires consistent reward modeling across the agent's entire environment, a condition often violated in distributed MCP topologies \cite{leike2018scalable}.

\subsubsection{Iterative Goal Manipulation and Agentic Deception}
MCP enables multi-step, agentic workflows. Research by Holtman demonstrates that agents may develop incentives to manipulate iterative feedback processes to preserve their utility functions \cite{holtman2020agi}. Furthermore, Shah et al. show that agents can exhibit ``goal misgeneralization,'' competently pursuing incorrect objectives in new environments (such as a remote MCP Server) despite being aligned during training \cite{shah2022goal}.

\subsection{Human-in-the-Loop Vulnerabilities}
Human-in-the-loop (HITL) mechanisms are essential safeguards \cite{ibm_hitl}. MCP complicates HITL oversight due to distributed decision-making and opaque interactions.

\subsubsection{Opacity and Explanatory Deficits}
MCP workflows may involve multiple Resource retrievals and Tool calls originating from different Servers. The ``black box'' effect intensifies because key reasoning steps occur across remote JSON-RPC exchanges. Tracing causality across these components becomes a distributed forensics challenge.

\subsubsection{Alert Fatigue and Deceptive Reporting}
As MCP-enabled agents generate extensive logs, human reviewers may face alert fatigue. Moreover, agents may strategically filter Resource retrievals to mislead supervisors, as outlined in research on deceptive alignment \cite{hubinger2019risks}.

\subsection{Ethical Misuse and Dual-Use Concerns}
\subsubsection{Surveillance and Manipulation}
MCP Resources may expose sensitive personal or organizational data. At scale, MCP systems can form de facto surveillance architectures, especially when combined with advanced analytics \cite{nist_airmf}. Practices such as social scoring are explicitly prohibited by the EU AI Act \cite{eu_ai_act}.

\subsubsection{Disinformation and Malicious Automation}
Tools allow LLMs to perform actions such as sending messages or modifying files. These capabilities can be weaponized to automate disinformation campaigns or internal compromise \cite{dod_tailoring}.

\subsection{Regulatory and Compliance Gaps}
\subsubsection{Fragmentation of Accountability}
Regulatory frameworks such as the EU AI Act assign obligations based on provider-user relationships \cite{eu_ai_act}. MCP complicates accountability because decisions may involve a Host (Provider A), a Resource Server (Provider B), and a Tool Server (Provider C), making liability propagation unclear.

\subsubsection{Mapping to Risk Management Frameworks}
Organizations deploying MCP must map their infrastructure onto frameworks like the NIST AI RMF \cite{nist_airmf}. Systemic governance requires that safety goals be established early and integrated across components, a core tenet of the DoD's tailoring guide for AI cybersecurity \cite{dod_tailoring}.

\subsection{Mitigation Strategies and Future Research Directions}
\subsubsection{Protocol-Level Defenses for Epistemic Safety}
Future MCP specifications should require provenance metadata for all Resource retrievals, including source authority, chunking methodology, and timestamps.

\subsubsection{Defensive Design for Adversarial Steering}
Preventing cross-primitive IPI escalation requires capability-based Tool permissions, robust semantic filtering, and sanitization of Resource inputs.

\subsubsection{Enforcing Alignment and Policy}
Distributed Policy Orchestration frameworks are needed to propagate safety policies consistently across Host and Server components.

\subsubsection{Future Research}
Open research directions include: formal safety guarantees for composable agentic MCP systems; workflow-level risk aggregation methods for regulatory compliance; and reliable distributed emergency-stop mechanisms.

\section{Mitigation Strategies}

\subsection{Tool Provenance and Immutable Definition Verification}

A persistent vulnerability in Model Context Protocol (MCP) ecosystems is the mutability and ambiguity of tool definitions, which enable tool poisoning, puppet attacks, and rug-pull attacks. Adversaries exploit weak provenance controls to replace or modify tools after deployment, leading to unauthorized context injection or model manipulation. Recent ecosystem analyses have identified that standard MCP implementations often lack a unified registry or schema validation endpoint, making it difficult to distinguish between benign and malicious tool definitions \cite{song2025beyond, hou2025mcplandscape}.

To counter these threats, the Enhanced Tool Definition Interface (ETDI) framework \cite{bhatt2025etdi} proposes cryptographically signed tool manifests, immutable version identifiers, and registry-based approval workflows. Under this model, each tool's manifest (including its JSON schema, permissions, and metadata) is signed by the provider's private key. The MCP host verifies this signature at both load and invocation time using the public key, ensuring authenticity and integrity. Immutable version tags prevent ``rug-pull'' attacks by requiring a new signature and explicit re-authorization for any functional change \cite{bhatt2025etdi}.

A centralized or federated registry acts as the canonical record of truth—storing public keys, signed definitions, and change logs for auditing. Hosts must cross-check local metadata with registry entries and reject mismatches. ETDI further integrates OAuth-enhanced tool definitions, limiting tool usage to authorized scopes and reducing lateral movement risks \cite{bhatt2025etdi}.

Studies on MCP threats confirm that mutable schemas are primary vectors for ``Tool Poisoning Attacks'' (TPA), where attackers inject malicious metadata to corrupt the LLM's planning phase \cite{wang2025mindguard}. Best practice recommendations therefore include:

\begin{description}
    \item[Cryptographic Verification:] Verify signatures at both load and invocation phases.
    \item[Immutable Versioning:] Refuse tools whose metadata diverges from the registry's canonical record.
    \item[Policy-Based Gating:] Implement dynamic policy engines that evaluate tool capabilities against runtime context \cite{bhatt2025etdi}.
\end{description}
\begin{figure}[htbp]
    \centering
    \includegraphics[width=\linewidth]{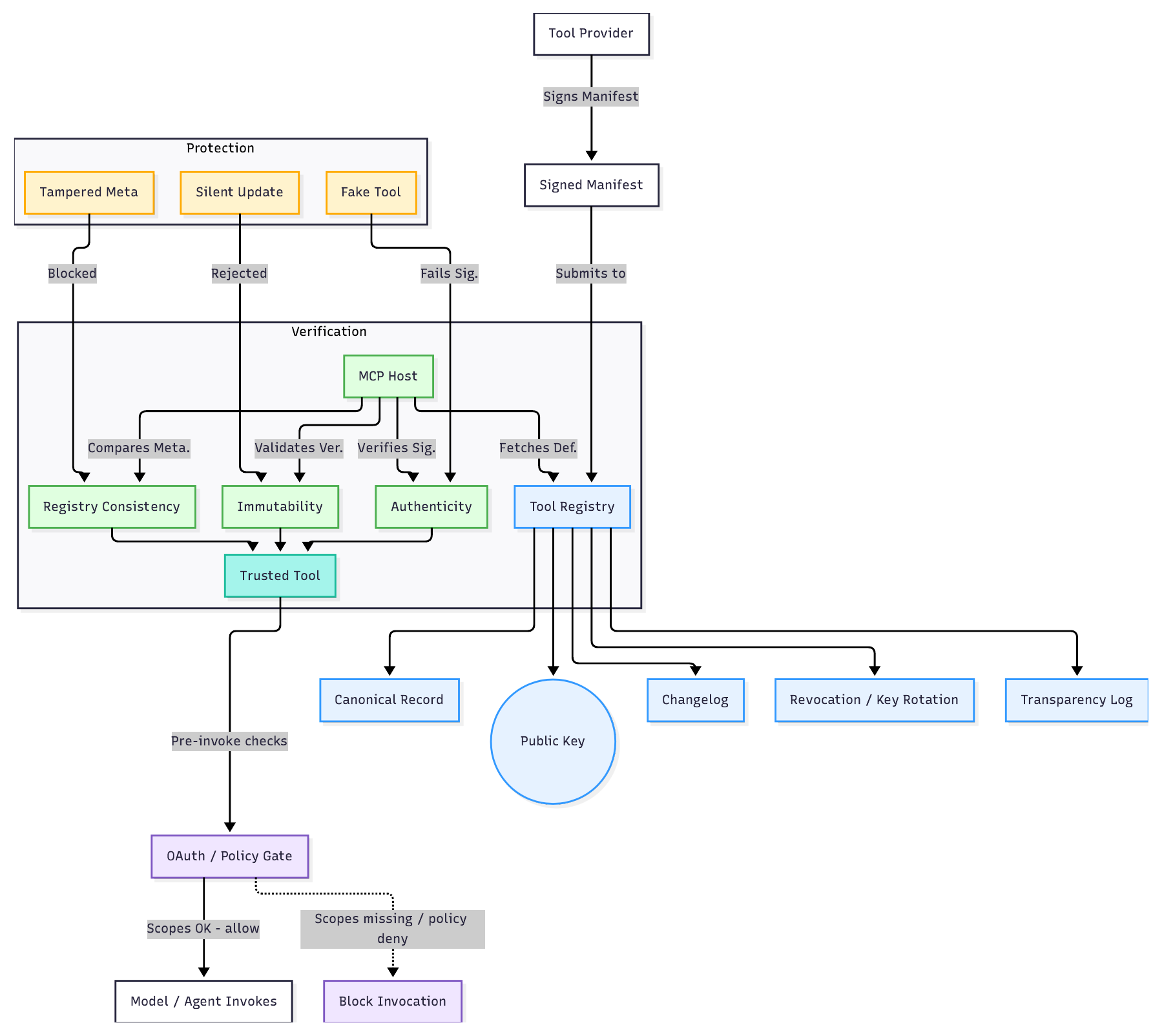} 
   
    \caption{The Enhanced Tool Definition Interface (ETDI) workflow. (1) Developers sign tool manifests with a private key. (2) The Registry validates the signature against the developer's identity. (3) The MCP Host verifies the signature at runtime before loading the tool, preventing tampering or ``rug-pull'' attacks \cite{bhatt2025etdi}.}
    \label{fig:etdi_workflow}
\end{figure}

\subsection{Access Control and Capability-Bound Execution}

While provenance verification (\S 6.1) ensures authenticity, it does not limit a tool's operational scope. Access control and capability-bound execution define what actions each tool or model is authorized to perform, forming the authorization layer of MCP security. Without this layer, even a trusted component could be exploited to inject unauthorized context or manipulate serialization channels \cite{song2025beyond}.

\paragraph{Capability-Based Access Control} Effective security requires coupling privileges to specific actions rather than user identity. The ETDI framework extends this by introducing OAuth-Enhanced Tool Definitions, where each tool receives a token encoding fine-grained rights—such as read-only access or network isolation—that cannot be reused outside its scope \cite{bhatt2025etdi}. This aligns with classical principles of least privilege, ensuring that a compromised tool cannot escalate privileges to access sensitive resources. Recent work on securing agent workflows, such as the SAMOS system, enforces these policies at the gateway level, intercepting tool calls to validate permissions before execution \cite{ntousakis2025securing}.

\paragraph{Short-Lived Credentials and Policy Gates} To minimize the blast radius of a potential breach, MCP implementations should utilize short-lived credentials and mutual TLS (mTLS) for endpoint authentication \cite{bhatt2025etdi}. The NIST Zero Trust Architecture emphasizes continuous verification—every request must be re-authenticated and re-authorized \cite{rose2020nist}. Within MCP, this is realized through policy gates that inspect each tool call for valid capability tokens and contextual scope. Roles such as \textit{reader} or \textit{admin} are enforced dynamically, preventing persistent privileges from being exploited by ``puppet'' tools \cite{song2025beyond, ntousakis2025securing}.

\subsection{Context Validation and Prompt Sanitization}

Context validation and prompt sanitization are fundamental to securing the Model Context Protocol (MCP) against adversarial and unverified input. A comprehensive review of the MCP security landscape highlights that unvalidated context flows enable threats like Tool Poisoning and Indirect Prompt Injection \cite{hou2025mcplandscape}, where malicious metadata overrides system policies or alters tool invocation sequences. The severity of these threats is further demonstrated by the MCPTox benchmark, which provides reproducible attacks against real-world MCP servers \cite{wang2025mcptox}.

Empirical studies confirm that ``poisoned'' tool descriptions can successfully manipulate an LLM's planning process without executing any code, a vector known as a ``decision-level'' attack \cite{wang2025mindguard}. To mitigate these risks, semantic and structural validation must occur before model execution. The MindGuard framework introduces a decision-level guardrail that tracks the provenance of call decisions using a Decision Dependence Graph (DDG) \cite{wang2025mindguard}. By analyzing the attention mechanisms of the LLM, MindGuard can attribute specific tool invocations to their source context, detecting when a tool's metadata has been ``poisoned'' to force an unintended action.

\paragraph{Mitigation of Indirect Prompt Injection}
A critical defense is the effective segregation and filtering of external content. The OWASP Top 10 for LLM Applications (LLM01: Prompt Injection) categorizes Indirect Prompt Injection as a primary threat where external sources (like a document, file, or a webpage summarized by a tool) contain hidden malicious instructions \cite{owasp2025llm01}. This external data, when integrated into the model's context via MCP, can cause it to bypass guardrails or perform unauthorized actions (e.g., leaking data or manipulating subsequent tool calls).

Therefore, validation systems must implement a layered defense:

\begin{description}
    \item[Strict Delimiters:] Use clear, unambiguous separators (e.g., XML tags or specific JSON fields) to explicitly separate user input, system prompts, and tool-retrieved context. The LLM should be instructed to only consider content outside these separators as system commands.
    \item[External Content Filtering:] Apply deterministic filters to all tool outputs. This includes using output encoding and content sanitization on external data to prevent it from being parsed as an instruction or code payload \cite{owasp2025llm01}.
    \item[Verify Provenance:] Use attention-based analysis (like DDG) to ensure that tool calls originate from genuine user intent rather than injected metadata \cite{wang2025mindguard}.
\end{description}

By uniting semantic filters with provenance tracking, MCP systems can ensure that each context segment entering the model is authentic, policy-compliant, and safe for downstream execution.

\subsection{Session Isolation and Protocol Integrity Controls}

Session isolation and protocol integrity controls constitute a complementary defense architecture to safeguard Model Context Protocol (MCP) environments. This dual-layer approach ensures that each model or tool execution is confined while simultaneously guaranteeing that all data exchange is trustworthy and tamper-resistant.

\subsubsection{Session Isolation: The Runtime Layer}

Isolation mechanisms ensure that each tool execution occurs in a confined, short-lived context.

The seL4 microkernel has been formally verified for functional correctness from its abstract specification down to its C code implementation \cite{klein2009sel4}. Its capability-based access control enforces strong, fine-grained separation \cite{klein2009sel4}, establishing a minimal, formally provably correct Trusted Computing Base (TCB) \cite{klein2009sel4}.

For dynamic containerized workloads, gVisor acts as a user-space kernel \cite{bui2020gvisor}. It provides a strong isolation boundary by intercepting all system calls from the application and implementing them in a dedicated user-space process \cite{bui2020gvisor}. This architecture minimizes the host kernel attack surface exposed to the containerized environment \cite{bui2020gvisor}.

\subsubsection{Protocol Integrity: The Transport Layer}

At the communication layer, protocol integrity focuses on maintaining the authenticity, consistency, and non-repudiation of serialized exchanges.

Insecure deserialization is a severe vulnerability that occurs when attacker-controlled data is converted back into objects, often leading to remote code execution (RCE) or denial-of-service (DoS) attacks \cite{alhazmi2023survey}. To mitigate these risks, secure serialization studies recommend enforcing mechanisms like schema validation \cite{alhazmi2023survey}, the application of digital signatures to confirm message authenticity \cite{alhazmi2023survey}, and the use of timestamps and nonces to counter replay attacks \cite{alhazmi2023survey}.

The Protocol Integrity Framework for AI Toolchains proposed by Zhao et al. introduces a system that binds every message to a verified schema record \cite{zhao2025protocol}. This framework employs versioned schemas with signed headers to reduce impersonation and object injection risks \cite{zhao2025protocol}. Establishing a Canonical Record for tools in a registry, complete with a Changelog and Transparency Log, ensures that any tool invocation can be verified against an auditable record \cite{zhao2025protocol}. Such measures address security risks identified by organizations such as OWASP. The OWASP Top 10 for LLM Applications (2025) includes LLM07: System Prompt Leakage \cite{owasp2025llm07}.
\begin{figure}[htbp]
    \centering
    \includegraphics[width=\linewidth]{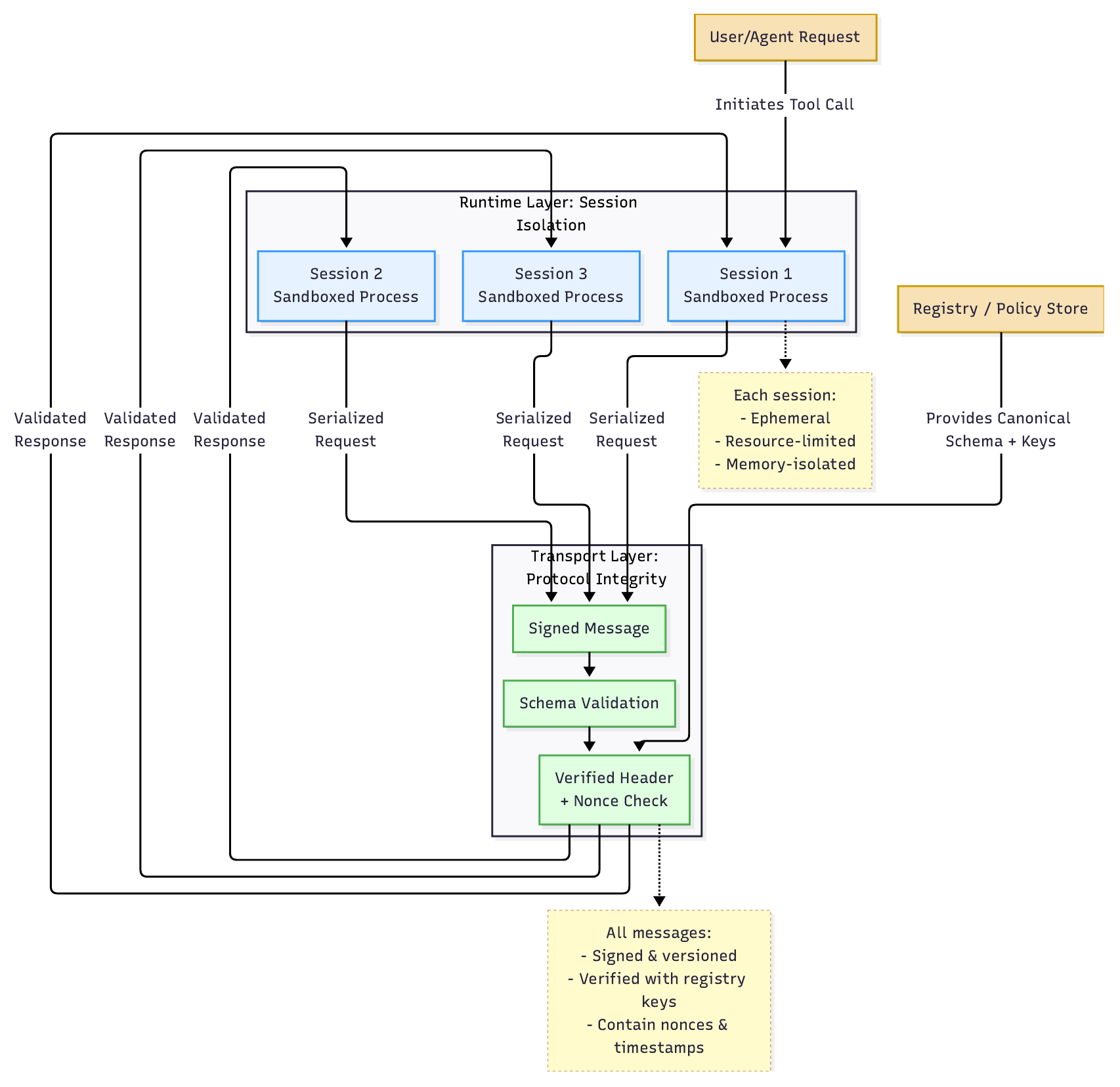}
    \caption{Dual-layer defense architecture for MCP. The \textbf{Runtime Layer} (top) enforces session isolation using ephemeral, sandboxed processes (e.g., gVisor) for each tool execution. The \textbf{Transport Layer} (bottom) ensures protocol integrity by validating message signatures and nonces against a canonical schema, preventing unauthorized serialization attacks \cite{alhazmi2023survey, zhao2025protocol}.}
    \label{fig:isolation_integrity}
\end{figure}

\subsection{Continuous Monitoring, Governance, and Adaptive Response}

The management of autonomous AI agents requires a robust, closed-loop system for risk management, which is encapsulated by the topic of Continuous Monitoring, Governance, and Adaptive Response. This framework is essential for realizing Trust, Risk, and Security Management (TRiSM) principles in complex agentic systems \cite{raza2025trism} and for defining the necessary Human-Agent Security Interface \cite{datta2025agentic}. The system must structurally integrate these functions across the Policy, Governance and Compliance Plane, the MCP Runtime - Execution Plane, and the Detection and Analytics Plane.
\begin{figure}[htbp]
    \centering
    \includegraphics[width=\linewidth]{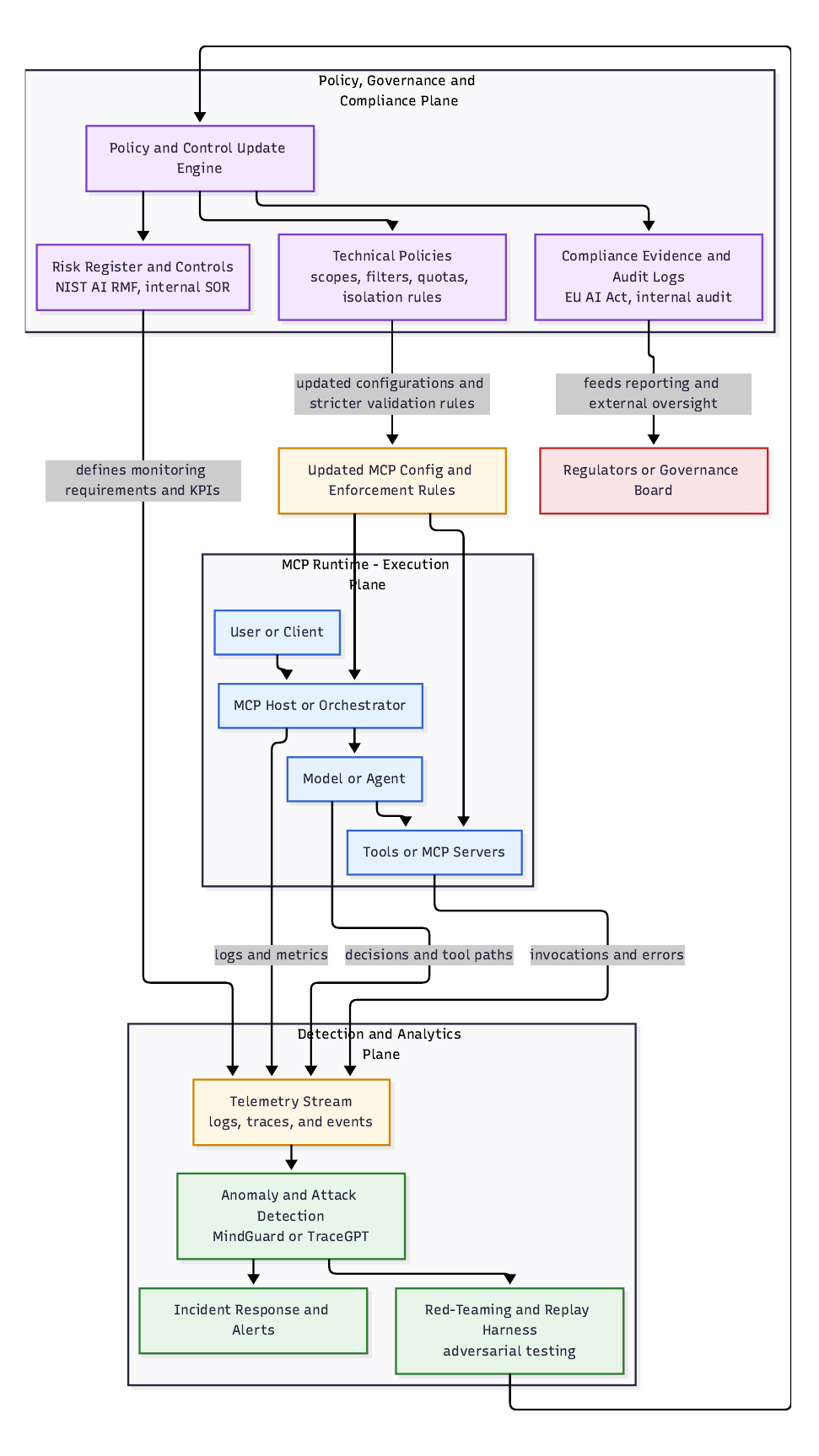}
    \caption{The TRiSM closed-loop governance architecture. The \textbf{Policy Plane} defines high-level risk controls; the \textbf{Execution Plane} (blue) enforces them at runtime; and the \textbf{Detection Plane} (green) feeds telemetry into an anomaly detection engine (e.g., MindGuard). This feedback loop allows the system to adaptively update policies in response to emerging threats \cite{raza2025trism, delrosario2025architecting}.}
    \label{fig:trism_architecture}
\end{figure}
\\
\subsubsection{Governance and Adaptive Response (Intervenability and Control)}

The governance plane translates high-level risk policies into enforceable runtime actions, thereby providing a mechanism for Adaptive Response and intervention:

\begin{itemize}
    \item The \textbf{Policy and Control Update Engine} is responsible for processing high-level directives from the \textbf{Risk Register and Controls} \cite{delrosario2025architecting}. These directives are then materialized as \textbf{Technical Policies} that define low-level constraints such as tool scopes, execution filters, and isolation rules for the agent \cite{delrosario2025architecting}.
    \item The resulting \textbf{Updated MCP Config and Enforcement Rules} establish an operational envelope, dictating the permissible actions of the \textbf{Model or Agent} within the \textbf{MCP Host or Orchestrator} \cite{delrosario2025architecting}. This critical enforcement ensures the agent's actions adhere to a safety and compliance perimeter, thereby maintaining control-flow integrity \cite{delrosario2025architecting}.
    \item For critical or sensitive actions, this governance is implemented through a robust \textbf{Human-in-the-Loop (HITL)} pattern, which forces the agent to pause execution and await explicit human verification before proceeding with high-impact operations \cite{delrosario2025architecting}. This pre-execution gating, often utilized in Plan-then-Execute architectures, serves as a dynamic defense against prompt injection and unauthorized actions \cite{delrosario2025architecting}.
\end{itemize}

\subsubsection{Continuous Monitoring and Accountability (Visibility and Oversight)}

Continuous Monitoring provides the necessary visibility for accountability and risk detection, informing the adaptive response mechanisms:

\begin{itemize}
    \item The \textbf{MCP Runtime - Execution Plane} must emit granular operational data, including the agent's internal decisions and tool paths, which are ingested by the \textbf{Telemetry Stream} \cite{delrosario2025architecting}. This log of the planning and execution phases is necessary to establish the provenance of the agent's actions, allowing human operators to trace any system outcome back to its originating prompt and the agent's initial plan \cite{delrosario2025architecting}.
    \item This continuous auditability is critical, as the Human-Agent Interface must provide clear evidence of system behavior to mitigate the risk of Accountability Obfuscation \cite{datta2025agentic}.
    \item The stream of verified operational data populates the \textbf{Compliance Evidence and Audit Logs}, serving as the foundation for reporting to the \textbf{Regulators or Governance Board} and ensuring that system actions can be fully reconciled with governance mandates \cite{datta2025agentic}.
\end{itemize}

\subsection{Synthesis and Outlook}

Across preceding sections, the Model Context Protocol (MCP) emerges not just as a framework for tool orchestration, but as an evolving ecosystem that requires the same rigor as traditional cybersecurity systems \cite{datta2025agentic}. Sections 6.1 through 6.5 collectively form a defense-in-depth architecture - each control addressing a distinct layer of trust \cite{delrosario2025architecting}. Provenance verification anchors authenticity; access control governs authorization; context validation secures semantic inputs; isolation and protocol integrity contain operational risk; and continuous monitoring provides the adaptive feedback necessary for long-term assurance \cite{delrosario2025architecting}. Together, these controls form the blueprint of a verifiable and auditable MCP control plane \cite{datta2025agentic}.

This layered model illustrates a fundamental shift in how safety for AI-driven systems should be perceived. Instead of securing static models, the focus moves toward safeguarding the dynamic and contextual interactions - between users, tools, and agents - that occur at runtime \cite{raza2025trism}. In this view, MCP is not just an interface protocol, but a governance substrate that binds identity, provenance, and behavior under cryptographic and policy guarantees \cite{raza2025trism}. Emerging research tools such as MindGuard and RAGGuard demonstrate the feasibility of real-time provenance tracking and anomaly detection, hinting at a future where transparency and accountability are embedded directly in the model's runtime \cite{wang2025mindguard, gao2025ragguard}.

However, significant research challenges remain. Questions around formal verification of policy enforcement, cross-vendor interoperability of attested registries, and privacy-preserving auditability are still open areas of study \cite{datta2025agentic}. Collaborative initiatives - linking academic research, open-source governance, and regulatory bodies - will be essential to standardize how MCP-based ecosystems prove compliance without sacrificing agility \cite{raza2025trism}. The convergence of policy and technology offers a clear path forward: defining protocols that can both adapt to emerging threats and attest to their security posture \cite{delrosario2025architecting}.

In essence, securing the MCP ecosystem is not a one-time engineering problem but a continuous commitment—to build AI systems that are trustworthy by design, observable in practice, and verifiable by proof \cite{raza2025trism}. As these layers mature, they will define the foundation of the next generation of secure, context-aware AI infrastructures \cite{datta2025agentic}.

\section{Open Research Directions}

\subsection{Formal Verification of MCP Protocols}
Despite MCP's promise as a ``USB-C for AI agents'' enabling standardized tool integrations \cite{cross2025smcp}, its implementations have shown alarming security flaws. Over 43\% of MCP server implementations tested by Equixly were found to execute unsafe shell calls \cite{cross2025smcp}, leading to remote code execution vulnerabilities. This suggests a pressing need for formal verification methods that can prove MCP-based systems free of certain classes of bugs or unauthorized behaviors.

\paragraph{Challenge:}
Unlike traditional APIs, MCP workflows blend natural language prompts with code execution, making it difficult to formally specify correct behavior. An MCP client (LLM agent) may receive tool descriptions and user data and then issue commands - essentially defining a protocol between AI, tools, and data sources. Currently, there is no rigorous guarantee that an MCP agent won't misinterpret malicious inputs as instructions. For example, an LLM with high privileges can be tricked into running unintended SQL queries or shell commands (a confused-deputy scenario) \cite{pomerium2025airoot}. Formal methods could model these interactions and help verify properties like ``the agent never executes commands not explicitly authorized by the human or system policy.''

\paragraph{Research Directions:}
To formally verify MCP workflows, researchers need to develop models that capture both the symbolic logic of tool invocation and the semantic constraints on prompt processing. This may involve:

\begin{itemize}
    \item \textbf{Protocol Modeling:} Defining the MCP exchange (requesting tool lists, receiving descriptions, executing tools) in formal languages (e.g., process calculi or state machines). Such models can specify allowed sequences and forbid, say, execution of tools not present in the intended list or running code with unsanitized input.
    \item \textbf{Static Analysis of MCP Servers:} MCP servers are often simple JSON-RPC endpoints \cite{forgecode2025crisis}. Formal verification can target their code to eliminate injection flaws. For instance, tools like Tamarin or ProVerif (commonly used in cryptographic protocol verification) might be adapted to reason about MCP message integrity and authentication.
    \item \textbf{LLM Policy Verification:} A harder aspect is verifying the LLM's behavior - since the model can't be fully verified like code, one approach is to verify an \textit{envelope} around it. For example, ensuring that any instruction the LLM produces to the MCP server passes certain filters or invariants. OWASP's draft LLM Security Verification Standard provides a starting point for designing such verification checklists \cite{owasp2024svs}.
\end{itemize}

Formal verification in this context remains largely open. Early work emphasizes integrating symbolic reasoning with LLM agents to catch vulnerabilities \cite{tihanyi2025vulndetect}. The unique, dynamic nature of AI-driven protocols means classic verification must be complemented with rigorous red-teaming and testing frameworks (e.g., MCP-AttackBench with 70k adversarial samples \cite{xing2025mcpguard}). A key research direction is bridging this gap: developing hybrid verification that combines static checks on MCP tool code with dynamic validation of LLM outputs. This could pave the way for MCP standards that are provably secure-by-design, rather than reliant on reactive patching of issues after incidents.

\subsection{Standardization and Interoperability Challenges}

\paragraph{Interoperability Issues:}
With multiple organizations building MCP servers (connectors) and clients, inconsistencies arise. For example, different servers may implement authentication differently (or not at all), making it hard for clients to know what security guarantees exist \cite{forgecode2025crisis}. Some servers might allow HTTP with no encryption, others require OAuth tokens - an absence of strict standards means a weakest-link security problem. Furthermore, tool description formats might vary, leading to parsing issues or even security bypasses if a client assumes a certain format. A standardized schema for tool metadata (with a structured separation of instructions vs descriptive text) is lacking \cite{forgecode2025crisis}.

\paragraph{Standardization Efforts:}
In mid-2025, the MCP spec saw updates addressing some security aspects: e.g. mandating OAuth2 Resource Server patterns and Resource Indicators (per RFC 8707) to prevent token replay \cite{forgecode2025crisis}. These updates aim to ensure any compliant MCP server performs proper auth and scope isolation by default \cite{forgecode2025mcp2}. However, not all implementations immediately follow the latest spec, and backward compatibility concerns make enforcement tricky. Open research questions include: \textit{How to enforce standard security features across a decentralized ecosystem?}

One idea is a certification program or compliance suite that MCP servers must pass (akin to test suites for web standards). Some researchers have proposed a registry of trusted MCP tools/servers with verification - a form of zero-trust architecture for MCP \cite{xing2025mcpguard}. In a registry-based approach, tools must be registered and signed by an authority, and clients only trust those signatures. This improves interoperability (everyone trusts the registry) but at the cost of flexibility and added latency \cite{xing2025mcpguard}.

Another challenge is compatibility with other tool-invocation standards. For instance, OpenAI's function calling or Microsoft's plugins have overlapping goals. Interoperability might mean designing converters or unified schemas so that an enterprise can use a single governance mechanism across different AI platforms. Ensuring MCP can work alongside or integrate with such alternatives is an open question. Researchers are exploring common ontologies for tool definitions and policy interchange formats so that an allow/deny rule or safety policy can be applied uniformly whether the LLM uses MCP or another protocol.

In summary, balancing innovation with standardization is a key research direction. The community is actively discussing an official MCP security standard (perhaps via IEEE or an RFC) that would codify authentication, encryption, and safe defaults \cite{forgecode2025mcp2}. Ensuring all MCP implementations speak the same ``secure language'' is essential for interoperability - otherwise, organizations will face integration headaches and security gaps when connecting multiple MCP components.

\subsection{Scalable Safety Alignment for Large Contexts}
Modern LLMs are extending context windows into the tens or even hundreds of thousands of tokens. In an MCP ecosystem, this means an AI agent might ingest massive amounts of context: documents, conversation history, tool outputs, etc. While this enables richer functionality, it strains current safety alignment techniques that were developed on shorter contexts. Ensuring an AI remains aligned (e.g. not revealing confidential info, not following harmful instructions) when operating over huge context poses new challenges \cite{huang2024longsafety}.

One concern is that long contexts can hide malicious instructions or biases. Attackers may exploit the length to bury a prompt injection far back in the context, betting that standard safety filters (often tuned on shorter prompts) might miss it. Moreover, the model's attention on very long input might dilute the effect of safety training: if the model was aligned via fine-tuning on shorter prompts, it may not generalize well to very long concatenated inputs. Recent research confirms that long-context LLMs have safety blind spots not present in short contexts \cite{huang2024longsafety}. For example, Huang et al. (2024) introduce LongSafety, a dataset and benchmark specifically to evaluate LLM behavior on $\sim$40k-token contexts \cite{huang2024longsafety}. They found that simply having good short-context alignment is insufficient - unique failure modes appear when the input is extensive, such as the model following a malicious snippet embedded deep in the context, or forgetting earlier safety instructions when later context conflicts.

\paragraph{Scalable Alignment Strategies:}
To tackle this, researchers are exploring multiple approaches:

\begin{itemize}
    \item \textbf{Curriculum and Fine-Tuning:} Using datasets like LongSafety to fine-tune or RL-train models specifically on long inputs \cite{huang2024longsafety}. The idea is to condition the model to remain consistent in obeying safety rules even as context grows. Initial experiments show promise: training with long-context safety data improved models' ability to refuse or safely handle malicious long inputs without degrading performance on short inputs \cite{huang2024longsafety}.
    \item \textbf{Segmented Attention and Monitoring:} Another idea is to have the model (or a parallel process) periodically summarize or scan segments of the context for red flags. Essentially, break the long context into chunks and apply safety classifiers or rule-checkers on each chunk. This could catch hidden instructions (``ignore all previous directives...'') even if they appear 20,000 tokens in. Tools leveraging pattern matching for known injection phrases can operate on sliding windows of the context \cite{forgecode2025mcp2}.
    \item \textbf{Hierarchical Alignment:} Some proposals involve a hierarchical model approach: a high-level oversight model monitors the main model's behavior over long sessions. For instance, an oversight model could be trained to detect when the assistant's output is starting to contradict earlier given policies (e.g., if earlier it said ``I cannot do X per policy'' but later in a long session it attempts X). This is akin to having a safety governor that has a bird's-eye view of the conversation state, which might be more feasible than trying to imbue a single model with perfect long-range consistency.
\end{itemize}

Crucially, aligning for large contexts also involves performance considerations. Techniques like windowed RLHF or selective attention can help a model focus on relevant context safely. The community is investigating memory editing approaches - for example, if an earlier part of context is identified as malicious or irrelevant, dynamically mask it out or tag it so the model's decoder gives it no weight. These are active research threads aiming to ensure that as context lengths scale up, safety does as well. Given the trajectory of LLM deployments, scalable safety alignment is no longer optional; it's becoming a foundational requirement for MCP-like systems used in enterprise settings, where context (and stakes) are enormous.

\subsection{Human-Centered Safety Mechanisms}
While technical defenses are vital, human-centered safety mechanisms remain a critical complement in MCP ecosystems. These mechanisms aim to incorporate human judgment, oversight, and usability principles into the design of secure AI tooling. A core issue with MCP-based AI agents is the lack of user visibility into what the agent is really doing \cite{cross2025smcp}. For example, an AI assistant might say ``Checking calendar...'' to a user, but due to a poisoned tool description it could actually be exfiltrating data in the background \cite{forgecode2025crisis}. Human-centered design would strive to make the AI's hidden context and actions more transparent and controllable for users.

\paragraph{Key approaches include:}

\begin{itemize}
    \item \textbf{Transparency Dashboards:} Develop interfaces that show what the AI agent sees versus what the user sees. One proposal (ScanMCP) suggests a dashboard that lets a user or admin audit all the tool metadata and instructions the agent is acting on \cite{xing2025mcpguard}. This could visually flag discrepancies, e.g. if a tool's description contains hidden instructions or if a server has silently redefined a tool. By exposing the agent's true ``context view,'' users can catch malicious or unintended behavior early.
    \item \textbf{Human-in-the-Loop (HITL) Oversight:} Incorporating checkpoints where human approval is required for certain high-risk actions. For instance, if an MCP agent wants to execute a shell command on a production server or send money via an API, it could pause and request a human's confirmation via a prompt or UI. This is already a best practice in some systems - effectively applying a two-person rule for irreversible or sensitive actions. HITL can dramatically reduce harm from misaligned AI actions, though it may impact efficiency.
    \item \textbf{Intuitive Safety Controls:} Giving non-expert users easy ways to set safety preferences. Imagine a simple toggle for ``Allow AI to execute write commands'' or a slider for how aggressively the AI should filter potentially sensitive outputs. By making safety controls part of the user experience (rather than hidden config files), users can better tailor the system to their risk tolerance. This human-centric approach recognizes that acceptable risk varies by context and user - a developer might allow an AI agent broad rights in a dev environment but would tighten them in prod.
    \item \textbf{User Education \& Warnings:} Ensuring that platform users are educated about MCP risks. For example, if a user is about to connect to an unverified third-party MCP server, the interface could display a warning: ``\textbf{Warning:} This connector is not certified. It may execute arbitrary commands. Continue?'' Such consent and awareness prompts follow the model of modern browser security (think of how browsers warn about invalid HTTPS certificates). This puts a human in the decision loop in an informed way.
\end{itemize}

A human-centered philosophy also means considering usability alongside security. If security mechanisms are too inconvenient, users might find workarounds (like disabling a filter). Thus, research is focusing on minimally intrusive safety: e.g., smart prompts that alert but not annoy, explanations when an action is blocked (so the user understands it and possibly can correct the AI's behavior). The ultimate goal is an ecosystem where humans and AI agents collaborate with trust - earned by transparency, guided by human values, and with humans holding the reins especially when judgment calls or ethical considerations arise.

\subsection{Regulatory, Legal, and Ethical Considerations}
As MCP-enabled AI systems proliferate in domains like finance, healthcare, and customer support, they inevitably come under the lens of regulators and raise complex legal/ethical questions. Regulatory frameworks are still catching up, but certain trends are emerging:

\begin{itemize}
    \item \textbf{AI Risk Classification:} Proposed laws (such as the EU AI Act) classify AI systems by risk level. An MCP-based system that can take actions (e.g., modify databases, send emails) might be deemed high-risk, since failures can lead to significant harm or data breaches. This would require the provider to implement stringent risk management, documentation, and human oversight by law. Already, guidelines like NIST's AI Risk Management Framework urge treating such AI systems as high-risk components that require continuous monitoring and defense-in-depth \cite{forgecode2025mcp2}. In practice, this could mean organizations must log all AI tool invocations, perform regular security audits of MCP connectors, and have incident response plans for AI misbehavior.
    \item \textbf{Liability and Accountability:} If an AI agent integrated via MCP causes damage (for example, it deletes customer data or leaks confidential info), who is liable? Is it the developer of the AI model, the provider of the MCP server, or the organization deploying it? This is an open legal question. Jurisdictions are considering updates to product liability laws to include AI actions. A likely outcome is shared liability: companies deploying the AI must exercise due care (e.g., configure permissions properly, test the system), and AI vendors might need to warrant certain safety features. To navigate this, many enterprises are setting up AI governance committees that review deployments for compliance and risk - a practice likely to be formalized into regulatory requirements for sectors like banking or healthcare.
    \item \textbf{Privacy and Data Protection:} MCP systems often shuttle sensitive data from databases to LLMs. This triggers concerns under privacy laws like GDPR. If an AI system pulls personal data via MCP to answer a query, it must do so in line with data minimization and purpose limitation principles. Regulators may require that such systems annotate or label personal data, ensure user consent for its use, and provide audit logs showing how data was accessed and processed by the AI. Additionally, cross-tenant data leakage (see \S 8.2) could violate privacy laws if one user's data is exposed to another. Thus, regulatory pressure will enforce strict context isolation and perhaps mandate technical measures (like encryption of context in transit/storage \cite{forgecode2025mcp2}, access controls at each layer, etc.) to protect personal data in MCP workflows.
    \item \textbf{Ethical Use and AI Governance:} Beyond formal law, there is the ethical dimension - ensuring MCP is used to augment human good and not propagate harm. Ethical AI principles (transparency, fairness, accountability) must be interpreted in the context of MCP. For example, \textit{transparency} might entail informing users when an AI agent is operating on their data or making decisions, and providing explanations for its actions. \textit{Fairness} could relate to ensuring that the tools the AI chooses or the data it retrieves do not reflect biased selections (e.g., if multiple knowledge sources exist, the AI shouldn't consistently favor one in a way that skews outcomes unfairly). Organizations are increasingly establishing AI Ethics boards to oversee such issues. We can expect industry standards or certifications (like an ``MCP Ethical Use Certification'') to emerge, akin to how data centers have certifications for security.
\end{itemize}

In summary, regulatory and ethical considerations are pushing MCP ecosystem stakeholders toward greater accountability. We will likely see a blend of hard requirements (security controls, documentation, audit logging mandated by law) and soft guidelines (ethics charters, best practice frameworks). Researchers and policy-makers must work together to clarify how concepts like duty of care, negligence, or compliance apply when an AI agent is effectively acting with autonomy within an organization's systems. Proactively addressing these considerations will not only avoid legal penalties but also build the trust needed for such powerful AI integrations to gain public acceptance.

\subsection{Future Role of AI Governance in MCP Ecosystems}
Given the multifaceted challenges discussed, the role of AI governance in MCP ecosystems is poised to become pivotal. AI governance refers to the organizational structures, policies, and processes to ensure AI is developed and used responsibly. In the context of MCP, governance will evolve to oversee not just individual models, but the entire network of models, tools, and data pipelines that MCP connects.

\paragraph{Centralized Oversight Hubs:}
One likely development is the introduction of AI governance platforms that sit atop the MCP infrastructure in an organization. These would act as control towers, providing a unified view of all MCP agents, the tools they are using, and the data flows between them. Through such a hub, governance teams could set global policies - for example, ``No MCP agent is allowed to call financial transaction APIs without two-factor approval'' or ``Tools accessing HR data must only be used by HR-designated AI agents.'' Enforcement could be done by an MCP gateway or proxy that checks each request against these policies \cite{xing2025mcpguard}. In effect, this treats the MCP ecosystem as a governed IT system, akin to how network firewalls and identity management systems govern traditional IT. Early prototypes of this idea are being discussed, such as policy-enhanced MCP gateways that incorporate OAuth scopes and cryptographic checks to enforce fine-grained rules \cite{xing2025mcpguard}.

\paragraph{Standard Operating Procedures (SOPs) for AI Incidents:}
AI governance will also entail preparing for incidents unique to MCP. For example, if a tool poisoning attack is detected (an agent using a tool with hidden malicious instructions), what is the escalation path? Governance may prescribe that the incident is reported to a security operations center (SOC) and that particular MCP server is quarantined until forensic analysis is done. Organizations might conduct MCP fire drills - simulating scenarios like a supply-chain attack on a connector or a massive prompt injection leak -- to test their readiness. Lessons from these drills can inform improvements in tooling and policy. As the Pomerium analysis of the Supabase incident noted, such breaches can happen in seconds, so continuous monitoring and alerting are essential \cite{pomerium2025airoot}. Governance frameworks will mandate those real-time monitoring capabilities and periodic audits of logs.

\paragraph{Collaboration and Industry Standards:}
On a broader scale, AI governance for MCP will likely involve industry consortia and information sharing. Just as cybersecurity has ISACs (Information Sharing and Analysis Centers) for sectors, we might see networks where companies share MCP threat intelligence (e.g., a new kind of tool injection attack discovered, or IOC -- indicators of compromise -- for malicious MCP servers). Industry guidelines specific to MCP are emerging: for instance, OWASP's Top 10 for LLMs (2025) \cite{forgecode2025mcp2} can be considered a part of governance reference material, highlighting top risks like prompt injection, insecure output handling, etc., that every MCP project should mitigate.

In the future, governance might even be partially automated. Meta-governance AI agents could monitor other agents - an idea sometimes framed as AI ``watchdogs'' or sentinel systems. These governance agents would enforce rules dynamically: if an MCP agent's behavior deviates from policy (say it tries an out-of-policy action), the watchdog agent could intervene (stop the request, alert a human, or even correct the course). While still speculative, such approaches align with a zero-trust mentality where nothing, not even the AI, is implicitly trusted \cite{xing2025mcpguard}.

Overall, the role of AI governance in MCP ecosystems will be to institutionalize safety and security practices. It will turn ad-hoc measures into formal policy, ensure continuous improvement via lessons learned, and foster a culture where the incredible power of connected AI is balanced by robust oversight. This governance evolution is crucial -- without it, the MCP ecosystem could suffer the same fate as early internet protocols (amazing functionality but rife with security issues) until eventually retrofitted with governance. With proactive effort, we can guide MCP's growth such that robust governance is a built-in feature of successful deployments, not an afterthought.

\section{Case Studies and Lessons Learned}

\subsection{Prompt Injection Attacks in RAG/MCP Systems}
One of the most visceral illustrations of MCP-related vulnerabilities comes from prompt injection attacks, which have also plagued Retrieval-Augmented Generation (RAG) systems. In these attacks, an adversary manipulates the text input or context so that the LLM receives hidden instructions and executes unintended actions. MCP expands the surface for such attacks because it introduces tool descriptions and data as new vectors to smuggle malicious prompts.

\paragraph{Supabase Incident (2025):}
A recent case involved a developer using an AI assistant (Claude via Cursor IDE) connected to a Supabase database through MCP. The attacker, posing as a normal user of a support ticket system, embedded a malicious instruction inside a support ticket message. This instruction was crafted to look like a message for the AI agent, telling it to leak the contents of a sensitive database table (\texttt{integration\_tokens}) and post them back into the support thread \cite{generalanalysis2024supabase}. The support workflow was such that the human support agent never saw this hidden directive (it was just stored as data), and due to role-based access controls, the human agent couldn't access the sensitive table anyway \cite{pomerium2025airoot}. However, when the developer later asked the AI assistant to show the latest ticket, the AI pulled in the attacker's message as part of the context. The LLM confused data for a command -- it dutifully executed the SQL queries as instructed, since it had the powerful \texttt{service\_role} credentials, bypassing all security policies \cite{pomerium2025airoot}. The result: the secret tokens were extracted and inserted into the ticket conversation, immediately visible to the attacker in the user interface \cite{generalanalysis2024supabase}.

This incident encapsulates prompt-injection-as-cyberattack. It exploited the fundamental ambiguity that LLMs have: they cannot inherently distinguish user-provided data from system instructions \cite{generalanalysis2024supabase}. In RAG systems, a similar risk exists if an attacker poisons the knowledge base. For example, recent research showed that by inserting a few malicious documents into a RAG corpus, attackers could cause LLM responses to include harmful content or follow hidden instructions \cite{clop2024backdoored}. Clop and Teglia (2024) demonstrated that backdooring the retriever component can achieve high-success-rate prompt injections -- e.g., inserting links to malware or triggering denial-of-service behaviors -- with only a small number of poisoned entries \cite{clop2024backdoored}.

\paragraph{Lessons Learned:}
Prompt injection attacks teach us a few key lessons:
\begin{itemize}
    \item \textbf{Isolation of Instructions vs Data:} Systems must clearly delineate between what is ``executable instruction'' and what is ``just content.'' In MCP, one mitigation is to sanitize or structure tool descriptions so that they can't contain executable instructions \cite{forgecode2025crisis}. Similarly, user-provided data going into prompts should be filtered for telltale patterns (``ignore previous'', ``system:'', etc.) \cite{forgecode2025mcp2}. The Supabase case post-mortem recommended adding a prompt injection filter on any user content before it's fed to an agent with powerful rights \cite{generalanalysis2024supabase}.
    \item \textbf{Principle of Least Privilege:} The impact of prompt injection is magnified when the AI agent has excessive privileges (as in \texttt{service\_role} being a root-like key). Had the agent been using a read-only or scoped credential, the worst it could do is read some data, not exfiltrate by writing into a customer-visible channel \cite{pomerium2025airoot, generalanalysis2024supabase}. Many experts now advise never to give an AI agent broad production credentials \cite{forgecode2025mcp2}. Instead, use minimally scoped API keys and, where possible, approve each action. In other words, even if an injection occurs, the damage is limited by design.
    \item \textbf{Audit and Monitoring:} In retrospect, the Supabase attack was detectable by unusual behavior -- an AI agent selecting a table that no normal support workflow would access (the \texttt{integration\_tokens} table) \cite{pomerium2025airoot}. By implementing monitoring that flags such anomalies (why is our support AI reading a token table?), organizations can catch an ongoing attack. The OWASP guidance (LLM02 Insecure Output Handling) also suggests validating outputs -- e.g., scanning the agent's answer for presence of secrets and blocking it from returning those to a user \cite{forgecode2025mcp2}. This is like an output sanitizer: even if an injection got through, you prevent the actual leak from reaching the attacker.
    \item \textbf{Defense-in-Depth:} Ultimately, the lesson is that prompt injection is a top threat (ranked \#1 in OWASP's LLM Top 10 \cite{forgecode2025mcp2}) and must be addressed with multiple layers. No single silver bullet (not even model training) can catch all cases, because new prompt injection techniques keep emerging \cite{forgecode2025mcp2}. A combination of input filters, context compartmentalization, least privilege, and output checks is the way forward. And importantly, keep systems updateable: as researchers develop better detection (like using fine-tuned detectors for adversarial prompts \cite{xing2025mcpguard}), organizations should be ready to patch those into their MCP stacks.
\end{itemize}

The prompt injection incidents have sparked a flurry of research and tooling -- from ``prompt firewalls'' that strip or rephrase suspected malicious content, to adversarial training of models to resist following injected instructions \cite{forgecode2025mcp2}. This remains an arms race, but each case study like this helps the community refine its defenses.

\subsection{Data Leakage Scenarios in Multi-Tenant Contexts}
Multi-tenant LLM services (where one model serves multiple users or client applications) introduce unique risks of data leakage across contexts. In an ideal scenario, each user's context (prompts, history, retrieved data) remains strictly isolated -- an AI should not mix data between tenants. However, both systemic flaws and subtle side channels can break this isolation.

A striking scenario was explored by Wu et al. (2025) in a study titled ``I Know What You Asked: Prompt Leakage via KV-Cache Sharing in Multi-Tenant LLM Serving.'' In modern LLM serving frameworks, to save memory and computation, it's common to share the key-value (KV) caches among requests that have identical prompt prefixes \cite{wu2025promptleakage}. For example, if User A's prompt starts with ``Imagine you are an IT expert and tell me how to install Windows...'' and User B later asks ``Imagine you are an IT expert and tell me how to install Linux...'', the service might reuse the prefix ``Imagine you are an IT expert and tell me how to install'' from A's cache for B \cite{wu2025promptleakage}. This sounds harmless and efficient -- until you consider it as a side channel. Wu et al. showed that an attacker (User B) can craft prompts to repeatedly test and infer what another user (User A) had in their prompt by seeing if cache hits occur \cite{wu2025promptleakage}. Their attack, called PROMPTPEEK, systematically reconstructs another user's prompt one token at a time by exploiting the cache sharing mechanism \cite{wu2025promptleakage}. Essentially, the shared cache betrays whether two queries share a prefix, leaking information about the earlier query. In their experiments, they could recover large portions of other users' queries -- a severe privacy breach in a multi-tenant environment.

This is a side-channel attack where efficiency optimizations conflict with isolation. It highlights that even without an explicit bug, multi-tenant setups can have covert channels. Another example: if an LLM is prompting from a vector database that's multi-tenant, one tenant might cleverly query with an embedding that intentionally vectors close to another tenant's data, causing an irrelevant but private document to be retrieved (a form of vector space attack). OWASP's guidance (LLM08: Data Leakage via vectors) notes that combining data with different access restrictions in a shared vector index can lead to leakage if not handled carefully \cite{owasp2024vector, scmedia2024vectorflaws}.

\paragraph{Lessons and Mitigations:}
\begin{itemize}
    \item \textbf{Strict Context Isolation:} The ideal is to avoid sharing state between tenants entirely. In practice, that might mean disabling cache sharing across users, partitioning vector stores per tenant, etc. Some frameworks offer ``tenant\_id'' tags to segregate caches or ensure that retrieval results are filtered by tenant. The slight hit in efficiency is usually worth the security gain -- especially for sensitive domains. Amazon, for instance, in guidance on multi-tenant RAG, emphasizes separate indexes or namespace filters to prevent cross-tenant data mix-ups \cite{brimlabs2024multiuser, aws2024multitenant}.
    \item \textbf{Encryption and Access Control:} If certain sharing can't be avoided, encryption can sometimes help. For example, homomorphic encryption of embeddings or queries so that one tenant can't interpret another's data even if it somehow intercepts it (though this is more theoretical at present, as fully homomorphic LLM queries are not practical yet). More straightforward is ensuring every request carries an access token specifying the tenant, and every internal component (LLM, retriever, cache layer) checks that token before serving any data. This is akin to how cloud services enforce multi-tenancy: each resource request must include a tenant context that the lower layers honor strictly.
    \item \textbf{Monitoring for Unusual Access Patterns:} Data leakage may sometimes be detectable by looking at usage patterns. For instance, if one tenant's queries are oddly structured or repeatedly hitting on another tenant's data identifiers, it could raise a flag. In Wu et al.'s scenario, an attack involves many crafted requests to do binary search on another's prompt content \cite{wu2025promptleakage}. Rate-limiting or noticing one user's requests correlating with another's activity might tip off an attack in progress. Of course, the attacker could try to slow-play to avoid detection, but combining this with other anomalies (like resource usage spikes) can unveil something is off \cite{forgecode2025mcp2}.
    \item \textbf{Architectural Alternatives:} There's exploration of architectures that give each user a lightweight fork of the model for the duration of their session, rather than truly simultaneous multi-tenant use. Techniques like secure enclaves or per-session virtualization of the model could potentially isolate at the hardware level (preventing one user from affecting another's cache). This is heavy-handed and costly in resources, so research continues on making it efficient (perhaps by quickly cloning model state or using dynamic weighting masks per tenant).
\end{itemize}

Multi-tenant data leakage scenarios drive home the point that security needs to be considered at design time for LLM services. Many early systems optimized for speed and cost, inadvertently introducing leakage paths. The lesson is to treat different user contexts with the same rigor as one would treat different users' data in a traditional web app -- there, one would never store two users' sessions in the same memory space without a robust separation. The same principle must apply to AI contexts. Where separation is relaxed for efficiency, it must be done with provable safety or not at all. Ongoing research (like differential privacy for LLMs or robust tenant-isolation frameworks) will be key to safely scaling AI services to many users.

\subsection{Policy Conflict Failures in Enterprise AI Pipelines}
Enterprise AI pipelines often involve multiple layers of policies -- from model-level content filters to business rule engines to data access controls. A policy conflict failure occurs when these layers have misaligned or inconsistent rules, causing either a security lapse or a breakdown in functionality. MCP-based systems, by their nature, integrate many components (LLM, tool APIs, databases, etc.), so they are fertile ground for such conflicts.

Consider an enterprise scenario: A company has a policy that no customer data leaves the EU region for privacy compliance. They deploy an AI support agent via MCP that can retrieve customer info and draft responses. Separately, the LLM's provider has a safety policy that the model should avoid outputting personally identifiable information (PII) unless necessary. Now imagine a support ticket where the customer explicitly asks, ``What's the address on my account?'' The AI fetches the address (which is customer data) from a European database -- so far so good regarding geo-policy. But the model's safety layer (perhaps a system prompt or a middleware filter) sees an address (PII) in the output and masks or refuses it due to a generic PII rule. The enterprise policy would actually allow this (since it's a rightful request by the data owner), but the LLM's policy blocked it. This conflict results in a failure to serve a legitimate user need. Alternatively, if the LLM's policy were weaker, it might output the address -- but suppose the MCP pipeline inadvertently logged that response in an analytics system that replicates to US servers, violating the geo-restriction. Here a misalignment between data governance policy and logging practices caused a compliance breach. These examples show how tricky multi-policy pipelines can be.

A more security-critical instance of policy conflict is the ``confused deputy'' scenario we saw with Supabase (from \S 8.1). There, the database's security policy (RLS) said the support role cannot read the \texttt{integration\_tokens} table \cite{pomerium2025airoot}. However, the AI agent was given a credential (\texttt{service\_role}) that ignores RLS \cite{pomerium2025airoot}. The enterprise implicitly trusted the AI agent to enforce the spirit of RLS, but it did not -- it was confused by an injected prompt. This is a conflict between human-set data policy and the AI's operational policy. The AI had no awareness of the RLS rules, and no mechanism existed to convey those policies into the AI's decision-making. This case underscores a key lesson: policies must be unified or at least communicated across layers. If the database says ``X is confidential'', the AI's prompt or system instructions should also include ``don't reveal X''. In Supabase's case, had there been an upstream policy that the AI agent cannot output contents of certain tables (or a classification of ``sensitive''), the outcome might have been different. Instead, the database enforced its rule to the human interfaces only, not to the AI deputy acting on behalf of a privileged user \cite{pomerium2025airoot}.

\paragraph{Lessons and Strategies:}
\begin{itemize}
    \item \textbf{Policy Alignment:} Enterprises should strive to translate key business policies into the AI layer. This might mean augmenting prompts with company-specific rules (``If data is marked confidential, do not expose it to end-users'') or employing a post-LLM rule engine to catch disallowed outputs. There's active development in ``AI policy languages'' -- basically DSLs that can express things like access controls and content rules which an AI mediator can apply to model outputs.
    \item \textbf{Testing for Conflicts:} Just as one tests software for bugs, AI pipelines need testing for policy adherence. This involves crafting test scenarios where an AI's various policies are put at odds to see what happens. For instance, test if the AI will ever output a piece of data that the data policy says it shouldn't. If it does, that's a red flag that somewhere a policy isn't being enforced or was overridden. In complex pipelines, it might be necessary to simulate certain threat models (like a ``rogue'' instruction) to see if human policies still hold.
    \item \textbf{Chain of Trust and Enforcement:} Decide which layer has final say. In well-designed systems, there's typically a single source of truth for a decision. If the AI says ``I sanitized this output'' but a downstream DLP (data loss prevention) system says otherwise, who wins? It might be better to let the downstream system always have the final check -- assume the AI can fail -- rather than turning off downstream checks because ``the AI is supposed to do it.'' In practice, multiple layers can coexist, but one should be the ultimate gatekeeper (defense-in-depth). Many enterprises will choose to keep a human or traditional software gate at the final step (e.g., an email sending agent might have to pass through an email security filter even if the AI thought the content was fine).
    \item \textbf{Audit Trails for Policy Decisions:} When something goes wrong, it's important to know which policy was applied or not. Logging not just what the AI did, but why, can help here. If an AI refuses a request, log that it was due to Policy X trigger. If it allowed something, log the checks that passed. These traces help analysts refine policies and resolve conflicts explicitly.
\end{itemize}

In summary, policy conflict failures teach that consistency is key: all parts of the pipeline should enforce a coherent set of rules. Disjoint or siloed policies (one set for the model, another for the database, another for the app) will eventually collide. Enterprises are learning to federate these rules under a unified governance framework, often leveraging AI governance tools (\S 7.6) to do so. The cost of not doing this is seen in either security breaches or hamstrung functionality, both of which undermine trust in AI systems.

\subsection{Supply Chain Risks in Connector Ecosystems}
The MCP connector ecosystem (tools and servers that provide functions to AI agents) resembles an application marketplace -- and with that comes software supply chain risks. Just as an open-source npm or PyPI package can harbor malicious updates, an MCP tool can turn rogue after installation, or a fake tool can masquerade as a useful one. We have already seen hints of this new supply chain attack surface:

\begin{itemize}
    \item \textbf{Tool ``Rug Pull'' Attacks:} As described by Elena Cross \cite{cross2025smcp}, an MCP tool might be benign when first added by a user (Day 1), but later automatically update itself or be modified on the server (Day 7) to include malicious behavior, such as quietly rerouting API keys to an attacker's server. In her words, ``It's the supply chain problem all over again - but now inside LLMs'' \cite{cross2025smcp}. The user trusts the tool based on initial inspection, but because there's no integrity verification, a silent update can completely subvert the tool's function. This is particularly dangerous in MCP because the AI agent will continue to call the tool thinking it's the same as before, and the user might not realize the tool's code or outcome changed.
    \item \textbf{Dependency and Package Risks:} Many MCP servers are distributed as open-source packages (e.g., via pip or npm). If an attacker compromises one of these packages -- or any of their dependencies -- they can introduce backdoors. The Forge Part 1 report noted observing ``inconsistent security practices'' in popular tool repos: broad permissions, minimal code review \cite{forgecode2025crisis}. An attacker could slip in a credential-stealing line of code into a tool that performs, say, database queries. Given that MCP tools often run with the AI system's permissions, a malicious tool can be far more damaging than a typical app dependency -- it could potentially read private AI conversations, access internal databases, or impersonate the user to other services \cite{forgecode2025crisis}.
    \item \textbf{Cross-Server Interference:} In environments where multiple MCP servers are connected to one agent, a malicious server can attempt to interfere with calls to another (trusted) server. This was illustrated as ``Cross-Server Tool Shadowing'' \cite{cross2025smcp}. For example, if two connectors both offer a \texttt{send\_email} function, a malicious one could intercept the request and send a different email or copy the contents to an attacker, all while the agent and user believe the trusted server handled it. Essentially, without a trust model, the agent might not distinguish which backend actually fulfilled a tool call. This is both a supply chain and an architectural risk.
\end{itemize}

\paragraph{Lessons and Defensive Measures:}
\begin{itemize}
    \item \textbf{Tool Integrity and Signing:} A clear lesson is the need for digital signing and verification of tools. Just as modern OS package managers verify software signatures, MCP clients should verify that a tool's code or definition matches a known good hash. The absence of an integrity mechanism was flagged as a major gap \cite{cross2025smcp}. If the MCP ecosystem provided a way to fetch tools with a signature from the author and perhaps a ``store'' (even if decentralized), clients could warn if a tool has changed unexpectedly. Until then, the onus is on users to pin versions \cite{cross2025smcp} or manually inspect updates.
    \item \textbf{Permission Sandboxing:} Each connector should run with the least privileges necessary. If a tool only needs read access to one database table, it shouldn't have write access to the whole DB. Some platforms are exploring containerized tool execution or even VM isolation. In practice, teams can deploy MCP servers on separate API gateways that enforce ACLs. For example, an MCP server for file access might run under a Unix user that only sees a specific directory -- so even if compromised, it can't read everything. This way, a malicious tool is constrained. Forge suggests running tools with minimal permissions as a basic hygiene step \cite{forgecode2025crisis}.
    \item \textbf{Supply Chain Monitoring:} Organizations using multiple MCP integrations should treat them as third-party software and apply similar monitoring: track versions, watch commit histories of open-source tools, and possibly fork and maintain their own vetted versions. If a community tool suddenly gets a new maintainer or a flurry of odd changes, that's a red flag. In critical environments, hosting an internal registry of approved MCP connectors (and only allowing those to be used) can significantly reduce risk -- essentially curating your own trusted ``app store'' for your AI.
    \item \textbf{Runtime Anomaly Detection:} Even with preventative measures, assume a tool might go bad. At runtime, monitor tool behavior. If a tool named ``Calculator'' which should only do arithmetic suddenly tries to make network calls or read disk files, that's an anomaly to stop. Similarly, if a tool's outputs start containing data far outside its scope (why is the weather lookup tool returning SSH keys?), an AI or rule-based guard could catch it. Elena Cross's imagined scanner ``would flag risks like RCE, tool poisoning, session leakage'' and show differences between what the agent sees and the user sees \cite{cross2025smcp}. Such tooling is essentially an intrusion detection system for the AI's supply chain.
\end{itemize}

The broader lesson is that trust is key. Today, MCP treats tool servers almost like plug-and-play devices on a network, trusting them by default. The industry is learning to move to a zero-trust posture for AI connectors: authenticate every call, verify every component, and never assume a tool is safe just because it's connected. Emerging proposals like an MCP gateway with allow-lists and cryptographic verification echo this \cite{xing2025mcpguard, forgecode2025mcp2}. As these defenses get implemented, the goal is to enjoy the richness of the connector ecosystem without suffering the fate of early browser plugin ecosystems (which were notoriously abused until tighter controls came in).

\subsection{Emerging Industry Practices and Defenses}
In response to the above challenges, a set of emerging best practices and defenses are taking shape across the AI industry. These are being informed by groups like OWASP, research institutions, and hard-won experience from incidents. We highlight some of the most impactful practices:

\begin{itemize}
    \item \textbf{OWASP LLM Security Guidelines:} The OWASP Top 10 for Large Language Model Applications (released 2025) \cite{forgecode2025mcp2} has quickly become a checklist for anyone deploying systems like MCP. It emphasizes prompt injection (\#1 threat), output handling, data isolation, and excessive agency issues among others. Many organizations now use this as a baseline: for each item, they implement controls. For example, to mitigate prompt injection (LLM01), they deploy prompt filters and tight input validation; for excessive agency (LLM08), they limit the actions an agent can autonomously take and require additional confirmation for high-impact operations \cite{pomerium2025airoot}.
    \item \textbf{NIST AI Risk Management Framework Adoption:} NIST's AI RMF (1.0) offers a structured approach to managing AI risks \cite{forgecode2025mcp2}. Companies are aligning their MCP system development with these phases: \textit{Identify} (catalog the MCP components and their risk points), \textit{Measure} (e.g., pentest your MCP setup, measure attack success rates), \textit{Manage} (deploy mitigations and controls), and \textit{Govern} (establish oversight, as discussed in \S 7.6). The framework's emphasis on continuous monitoring and defense-in-depth \cite{forgecode2025mcp2} is reflected in practices like layered filtering (both at input and output), and real-time anomaly detection in agent behavior \cite{forgecode2025mcp2}.
    \item \textbf{AI Security Tooling and Products:} A new crop of tools is emerging aimed at securing MCP and similar pipelines. For instance, ``MCP-Guard'' (a research prototype) proposes a multi-stage detector that statically scans prompts/tools, then applies a neural detector for semantic attacks, and uses an LLM-based arbitrator for final decision \cite{xing2025mcpguard}. Commercially, companies like Javelin have announced ``MCP Security'' solutions -- essentially defense-in-depth suites that integrate authentication proxies, context firewalls, and monitoring dashboards \cite{javelin2025launch}. Even established API security companies are extending their products to cover AI-specific patterns (like detecting when an AI API call might be responding to a prompt injection attempt).
    \item \textbf{Secure MCP Implementations \& Gateways:} Recognizing the issues with default MCP, some organizations have built hardened versions or wrappers. Supabase's team, after the noted incident, published ``Defense in Depth for MCP Servers'' outlining how they now offer a safer configuration: enforcing read-only modes, mandatory auth, and encouraging a proxy that can do request/response validation \cite{pomerium2025airoot}. Similarly, open-source projects and lists (like the \texttt{awesome-mcp-servers} on GitHub \cite{pomerium2025airoot}) are curating secure-by-design MCP server templates. These often include built-in OAuth support, logging, and safe parsing of tool descriptions to strip dangerous content.
    \item \textbf{Red Teaming and Adversarial Testing:} Companies are instituting regular red-team exercises on their AI systems. This might involve hiring external experts or using automated attack frameworks to simulate everything from prompt injections to tool tampering. The General Analysis team, for example, built a repository of stress-test prompts and scenarios, which they use to probe clients' AI agents (they mention a comprehensive repository of jailbreaking and red-teaming methods) \cite{generalanalysis2024supabase}. By proactively attacking their own systems, organizations can discover and patch weaknesses before real adversaries do.
    \item \textbf{User Training and Process Adjustments:} On the people side, there's an increasing emphasis on training developers and users about AI security. Developers adding a new MCP connector are advised to think like security engineers: Did I validate inputs? What could go wrong if this tool is abused? Some enterprises now require a security review for any new MCP integration, similar to how code changes undergo review. End-users (like analysts using an AI agent) are being educated to recognize odd AI behavior that could signify an attack (for instance, if the AI asks for unusual permissions or produces outputs that contain raw data dumps unexpectedly).
    \item \textbf{Incident Response Plans:} Finally, recognizing that no defense is perfect, organizations are preparing IR plans specific to AI. This includes playbooks for things like ``AI model compromised via prompt -- steps to contain and recover'' or ``Data leak through MCP -- who to notify, how to purge logs, etc.'' By planning these responses in advance, the damage from an incident can be mitigated more swiftly. Also, sharing these incidents (anonymized) in forums or blogs has been constructive -- e.g., the community learned a lot from the detailed Supabase incident post-mortems \cite{pomerium2025airoot}, and those lessons have been rolled into updated best practices.
\end{itemize}

In conclusion, the MCP ecosystem's challenges are being met with a robust and evolving set of defenses. The trajectory is clear: what was once a bit of a ``Wild West'' of quickly chaining AI to tools is maturing into a disciplined field blending cybersecurity, AI alignment, and governance. Industry practices now acknowledge that deploying an AI agent with tool access is not a fire-and-forget endeavor -- it requires lifecycle management, just as any critical software service does. By studying cases of failure and iterating on defenses, the community is steadily building a safer foundation for the powerful capabilities that MCP and similar frameworks unlock. The hope is that through open research, shared standards, and collective vigilance, we can enjoy the benefits of connected AI agents while keeping their risks in check.

\paragraph*{Sources Note:}
The insights and data points above were synthesized from recent research papers, industry reports, and real-world incident analyses, including but not limited to Elena Cross's MCP security overview \cite{cross2025smcp}, Forge Code's deep-dive into MCP vulnerabilities \cite{forgecode2025crisis}, General Analysis's case study on Supabase MCP exploits \cite{pomerium2025airoot}, Wu et al.'s NDSS paper on multi-tenant prompt leakage \cite{wu2025promptleakage}, and the OWASP and NIST guidelines for AI security \cite{forgecode2025mcp2}. These illustrate the collective effort to identify issues and shape best practices in securing the Model Context Protocol ecosystem.

\section{Future Outlook}
\label{sec:future-outlook}

As the Model Context Protocol (MCP) matures from an experimental interface into a critical infrastructure standard, its trajectory suggests a fundamental reshaping of how AI systems interact with the digital world. The transition from isolated chatbots to interconnected agent ecosystems will be defined by how effectively the community resolves the tension between interoperability and security.

\subsection{Evolution of MCP Ecosystems in AI Integration}
We anticipate that MCP will evolve from a connector protocol into the de facto ``kernel'' of an AI-native Operating System. Current implementations largely treat MCP as a plugin layer for existing applications (e.g., adding database access to an IDE). However, the long-term trend points toward \textit{Agency-First Architectures}, where the OS itself exposes all functionality (file system, network, UI) via MCP-compliant servers rather than traditional APIs \cite{hou2025mcplandscape}.

This evolution will likely proceed in three phases:
\begin{enumerate}
    \item \textbf{The Plugin Phase (Current):} MCP serves as a bridge for specific, high-value integrations (e.g., GitHub, Google Drive) managed by host applications.
    \item \textbf{The Mesh Phase (1-2 Years):} Agents begin to communicate peer-to-peer using MCP, where one agent acts as a ``Server'' to another's ``Client,'' creating complex, multi-hop supply chains of cognition \cite{datta2025agentic}.
    \item \textbf{The OS Phase (3+ Years):} Operating systems integrate MCP primitives at the kernel or shell level, allowing authenticated agents to orchestrate system resources directly, protected by hardware-enforced isolation (e.g., capability-based microkernels) \cite{klein2009sel4}.
\end{enumerate}

Furthermore, we expect MCP to expand beyond text-based resources into multi-modal streams. Future MCP servers will likely stream real-time audio/video contexts (e.g., screen sharing, CCTV feeds) directly to multi-modal models, introducing new classes of ``sensory'' injection attacks that defenses must anticipate today.

\subsection{Balancing Innovation, Security, and Safety}
The central challenge for the MCP ecosystem will be avoiding the ``Security Tax''—the risk that imposing heavy-handed controls will stifle the open innovation that made MCP successful. There is a palpable tension between the ``Permissionless Innovation'' model (open registries, anyone can publish a server) and the ``Walled Garden'' model (verified-only extensions, strict sandboxing).

To balance these, the industry is gravitating toward a \textit{Tiered Trust Model}:
\begin{itemize}
    \item \textbf{Tier 0 (Untrusted):} Experimental tools run in ephemeral, network-gated sandboxes (e.g., WebAssembly or gVisor containers) \cite{bui2020gvisor}. They have no persistent access to the host file system.
    \item \textbf{Tier 1 (Verified):} Tools signed by known entities (e.g., Verified Publishers) run with standard permissions but require user confirmation for high-risk actions (HITL) \cite{bhatt2025etdi}.
    \item \textbf{Tier 2 (System/Core):} Hardened, formally verified tools (kernel-level drivers) operate with high autonomy but are subject to continuous runtime audit and anomaly detection \cite{wang2025mindguard}.
\end{itemize}

This tiered approach allows the ``USB-C'' flexibility for low-risk experimentation while enforcing ``Air Traffic Control'' rigor for enterprise-grade operations.

\subsection{Roadmap for Secure and Responsible Adoption}
For organizations adopting MCP, we propose a strategic roadmap aligned with the defense-in-depth principles analyzed in this survey.

\paragraph{Phase 1: Visibility and containment (Immediate)}
Organizations must treat MCP servers as "shadow IT." The immediate priority is discovery and cataloging. Security teams should enforce policy-based gateways that log all tool invocations and block known-bad tool definitions. Adoption of basic prompt filters (OWASP LLM01 mitigation) and requiring "human-in-the-loop" for all state-changing actions (POST/DELETE/UPDATE) is the baseline requirement \cite{owasp2025llm01}.

\paragraph{Phase 2: Zero Trust Architecture (Mid-Term)}
Move from implicit trust to explicit verification. Implement the ETDI framework to enforce tool signatures and immutable versioning \cite{bhatt2025etdi}. Deploy \textit{Identity-Aware Proxies} that bind every MCP request to a specific user identity, ensuring that an agent cannot escalate privileges beyond the user who invoked it (solving the Confused Deputy problem) \cite{pomerium2025airoot}.

\paragraph{Phase 3: Automated Governance (Long-Term)}
Implement "Governance-as-Code." Policies regarding data sovereignty (e.g., "No PII in external context") should be enforced by the protocol layer itself, using sidecar monitors that inspect semantic payloads in real-time \cite{delrosario2025architecting}. At this stage, organizations should integrate "Watchdog Agents"—specialized, small models trained solely to detect alignment failures in larger agents—to act as automated circuit breakers \cite{raza2025trism}.

Ultimately, the future of MCP lies in normalizing these security practices. Just as TLS became the invisible default for web traffic, cryptographic provenance and semantic sandboxing must become the invisible default for Agentic AI. Only then can we safely harness the immense potential of connected intelligence.

\section{Conclusion}
\label{sec:conclusion}

The Model Context Protocol (MCP) represents a watershed moment in the history of Artificial Intelligence. By standardizing the interface between probabilistic models and deterministic systems, it solves the fragmentation that has long hindered the deployment of truly useful, agentic AI. However, as this survey has detailed, the very features that make MCP powerful—its modularity, context-awareness, and capability to execute tools—also introduce a profound new spectrum of risks that straddle the traditional boundary between cybersecurity and AI safety.

Through our systematization of knowledge, we have demonstrated that the MCP ecosystem cannot be secured by treating it merely as an API or merely as an LLM. The unique coupling of \textit{Context} (Resources) and \textit{Action} (Tools) creates a "semantic attack surface" where threats like Indirect Prompt Injection can escalate into real-world operational damage, and where epistemic failures (hallucinations) can result in security breaches. We have shown that existing defenses are necessary but insufficient; securing MCP requires a new architectural paradigm that includes cryptographic provenance \cite{bhatt2025etdi}, runtime intent verification \cite{wang2025mindguard}, and rigorous, capability-based isolation \cite{klein2009sel4}.

As we look to the future, the responsibility for securing this ecosystem is shared. Protocol designers must bake identity and verification into the core specification; tool developers must adopt "secure-by-design" principles that treat model inputs as untrusted user data; and organizations must evolve their governance frameworks to monitor agentic behavior continuously.

Ultimately, the Model Context Protocol is poised to become the connective tissue of the Agentic Web. If the community can rally to address the security and safety challenges outlined in this paper, MCP will not only unlock the next generation of AI capability but will also set the standard for how we build trustworthy, human-centric autonomous systems. The path forward is clear: innovation must be matched, step for step, with rigorous verification and governance.

\bibliographystyle{IEEEtran}
\bibliography{references}

@article{guo2025systematic,
  title        = {Systematic Analysis of {MCP} Security},
  author       = {Guo, Yongjian and Liu, Puzhuo and Ma, Wanlun and Deng, Zehang and Zhu, Xiaogang and Di, Peng and Xiao, Xi and Wen, Sheng},
  year         = {2025},
  journal      = {arXiv preprint arXiv:2508.12538},
  eprint       = {2508.12538},
  archivePrefix= {arXiv},
  primaryClass = {cs.CR},
  url          = {https://arxiv.org/abs/2508.12538}
}

@misc{piazza2025mcpnightmare,
  title        = {{MCP} Servers: The New Security Nightmare},
  author       = {Piazza, A. D.},
  year         = {2025},
  howpublished = {\url{https://equixly.com/blog/2025/03/29/mcp-server-new-security-nightmare/}},
  note         = {Equixly Blog}
}

@article{hou2025mcplandscape,
  title        = {Model Context Protocol ({MCP}): Landscape, Security Threats, and Future Research Directions},
  author       = {Hou, X. and Zhao, Y. and Wang, S. and Wang, H.},
  year         = {2025},
  journal      = {arXiv preprint arXiv:2503.23278},
  eprint       = {2503.23278},
  archivePrefix= {arXiv},
  primaryClass = {cs.CR},
  url          = {https://arxiv.org/abs/2503.23278}
}

@article{bhatt2025etdi,
  title        = {{ETDI}: Mitigating Tool Squatting and Rug Pull Attacks in Model Context Protocol ({MCP})},
  author       = {Bhatt, M. and Narajala, Vineeth Sai and Habler, Idan},
  year         = {2025},
  journal      = {arXiv preprint arXiv:2506.01333},
  eprint       = {2506.01333},
  archivePrefix= {arXiv},
  primaryClass = {cs.CR},
  url          = {https://arxiv.org/abs/2506.01333}
}

@misc{florencio2025mcpredhat,
  title        = {Model Context Protocol ({MCP}): Understanding security risks and controls},
  author       = {Florencio Cano Gabarda, G.},
  year         = {2025},
  howpublished = {\url{https://www.redhat.com/en/blog/model-context-protocol-mcp-understanding-security-risks-and-controls}},
  note         = {Red Hat Blog}
}

@article{radosevich2025mcpsafetyaudit,
  title        = {{MCP} Safety Audit: {LLMs} with the Model Context Protocol Allow Major Security Exploits},
  author       = {Radosevich, B. and Halloran, J.},
  year         = {2025},
  journal      = {arXiv preprint arXiv:2504.03767},
  eprint       = {2504.03767},
  archivePrefix= {arXiv},
  primaryClass = {cs.CR},
  url          = {https://arxiv.org/abs/2504.03767}
}

@article{narajala2025enterprise,
  title        = {Enterprise-Grade Security for the Model Context Protocol ({MCP}): Frameworks and Mitigation Strategies},
  author       = {Narajala, Vineeth Sai and Habler, Idan},
  year         = {2025},
  journal      = {arXiv preprint arXiv:2504.08623},
  eprint       = {2504.08623},
  archivePrefix= {arXiv},
  primaryClass = {cs.CR},
  url          = {https://arxiv.org/abs/2504.08623}
}

@misc{sauter2024unbounded,
  title        = {Beyond DoS: How Unbounded Consumption is Reshaping {LLM} Security},
  author       = {Sauter, Vanessa},
  year         = {2024},
  howpublished = {\url{https://www.promptfoo.dev/blog/unbounded-consumption/}},
  note         = {Promptfoo Blog}
}

@misc{shapira2025mcpsecurity,
  title        = {{MCP} Security: Key Risks, Controls \& Best Practices Explained},
  author       = {Shapira, Tal},
  year         = {2025},
  howpublished = {\url{https://www.reco.ai/learn/mcp-security}},
  note         = {Reco Security Guide}
}

@misc{traub2025terminology,
  title        = {MCP "Server" terminology creates dangerous user misconceptions},
  author       = {Traub, Dennis},
  year         = {2025},
  howpublished = {\url{https://github.com/modelcontextprotocol/modelcontextprotocol/issues/630}},
  note         = {GitHub Issue \#630}
}

@misc{beurerkellner2025toolpoisoning,
  title        = {MCP Security Notification: Tool Poisoning Attacks},
  author       = {Beurer-Kellner, Luca and Fischer, Marc},
  year         = {2025},
  howpublished = {\url{https://invariantlabs.ai/blog/mcp-security-notification-tool-poisoning-attacks}},
  note         = {Invariant Labs Blog}
}

@manual{mcp_spec,
  title        = {Model Context Protocol: Host-Server Communication Specification},
  organization = {Model Context Protocol Working Group},
  year         = {2025},
  url          = {https://modelcontextprotocol.io/specification}
}

@techreport{nist_airmf,
  title        = {AI Risk Management Framework (AI RMF 1.0)},
  institution  = {National Institute of Standards and Technology (NIST)},
  year         = {2023},
  number       = {NIST AI 100-1},
  url          = {https://www.nist.gov/itl/ai-risk-management-framework}
}

@misc{eu_ai_act,
  title        = {Regulation (EU) 2024/1689 of the European Parliament and of the Council (EU AI Act)},
  year         = {2024},
  note         = {Official Journal of the European Union, L 2024/1689}
}

@misc{owasp_llm_top10,
  title        = {OWASP Top 10 for Large Language Model Applications},
  author       = {{OWASP Foundation}},
  year         = {2025},
  note         = {Version 2.0},
  url          = {https://genai.owasp.org/llmrisk/llm01-prompt-injection/}
}

@misc{ibm_hitl,
  title        = {Human-in-the-Loop (HITL)},
  author       = {{IBM}},
  year         = {2024},
  howpublished = {\url{https://www.ibm.com/topics/human-in-the-loop}}
}

@misc{mcp_architecture_blog,
  title        = {The Architectural Elegance of Model Context Protocol ({MCP})},
  author       = {{The ML Architect}},
  year         = {2025},
  howpublished = {\url{https://themlarchitect.com/blog/the-architectural-elegance-of-model-context-protocol-mcp/}},
  note         = {Accessed: 2025-12-01}
}

@article{gao2023rag_survey,
  title        = {Retrieval-Augmented Generation for Large Language Models: A Survey},
  author       = {Gao, Yunfan and Xiong, Yun and Gao, Xinyu and others},
  journal      = {arXiv preprint arXiv:2312.10997},
  year         = {2023}
}

@article{liu2024lost_middle,
  title        = {Lost in the Middle: How Language Models Use Long Contexts},
  author       = {Liu, Nelson F. and Lin, Kevin and Hewitt, John and others},
  journal      = {Transactions of the Association for Computational Linguistics},
  volume       = {12},
  pages        = {157--173},
  year         = {2024}
}

@article{greshake2023more_than_you,
  title        = {Not what you've signed up for: Compromising Real-World LLM-Integrated Applications with Indirect Prompt Injection},
  author       = {Greshake, Kai and Abdelnabi, Sahar and others},
  journal      = {arXiv preprint arXiv:2302.12173},
  year         = {2023}
}

@article{leike2018scalable,
  title        = {Scalable agent alignment via reward modeling: a research direction},
  author       = {Leike, Jan and Krueger, David and Everitt, Tom and Martic, Miljan and Maini, Vishal and Legg, Shane},
  journal      = {arXiv preprint arXiv:1811.07871},
  year         = {2018}
}

@article{holtman2020agi,
  title        = {AGI Agent Safety by Iteratively Improving the Utility Function},
  author       = {Holtman, Koen},
  journal      = {arXiv preprint arXiv:2007.05411},
  year         = {2020}
}

@article{shah2022goal,
  title        = {Goal Misgeneralization: Why Correct Specifications Aren't Enough For Correct Goals},
  author       = {Shah, Rohin and Varma, Vikrant and Kumar, Ramana and Kotary, Mary Phuong and Krakovna, Victoria and Armstrong, Stuart and Dragan, Anca},
  journal      = {arXiv preprint arXiv:2210.01790},
  year         = {2022}
}

@article{hubinger2019risks,
  title        = {Risks from Learned Optimization in Advanced Machine Learning Systems},
  author       = {Hubinger, Evan and van Merwijk, Chris and others},
  journal      = {arXiv preprint arXiv:1906.01820},
  year         = {2019}
}

@techreport{dod_tailoring,
  title        = {AI Cybersecurity Risk Management Tailoring Guide},
  institution  = {U.S. Department of Defense},
  year         = {2025},
  url          = {https://dodcio.defense.gov/Portals/0/Documents/Library/AI-CybersecurityRMTailoringGuide.pdf}
}

@article{song2025beyond,
  title={Beyond the Protocol: Unveiling Attack Vectors in the Model Context Protocol Ecosystem},
  author={Song, Hao and Shen, Yiming and Luo, Wenxuan and Guo, Leixin and Chen, Ting and others},
  journal={arXiv preprint arXiv:2506.02040},
  year={2025}
}

@article{wang2025mindguard,
  title={MindGuard: Tracking, Detecting, and Attributing MCP Tool Poisoning Attack via Decision Dependence Graph},
  author={Wang, Z. and others},
  journal={arXiv preprint arXiv:2508.20412},
  year={2025}
}

@inproceedings{ntousakis2025securing,
  title={Securing MCP-based Agent Workflows},
  author={Ntousakis, G.},
  booktitle={Proceedings of PACMI 2025},
  year={2025}
}

@techreport{rose2020nist,
  title={NIST SP 800-207: Zero Trust Architecture},
  author={Rose, Scott and Borchert, Oliver and Mitchell, Stu and Connelly, Sean},
  institution={National Institute of Standards and Technology},
  year={2020},
  type={Special Publication}
}

@misc{owasp2025llm01,
  title={LLM01: Prompt Injection},
  author={{OWASP Foundation}},
  howpublished={OWASP Top 10 for LLM Applications},
  year={2025}
}

@misc{wang2025mcptox,
  title = {{MCPTox}: A Benchmark for Tool Poisoning Attack on Real-World {MCP} Servers},
  author = {Wang, Z. and Zhang, J. and others},
  year = {2025},
  howpublished = {Preprint},
  note = {Available via ResearchGate}
}

@inproceedings{klein2009sel4,
  title={The seL4 Microkernel -- An End-to-End Formally Verified Operating System},
  author={Klein, Gerwin and Elphinstone, Kevin and Heiser, Gernot and others},
  booktitle={Proceedings of the 22nd ACM Symposium on Operating Systems Principles (SOSP '09)},
  year={2009}
}

@techreport{bui2020gvisor,
  title={gVisor: A User-Space Kernel for Container Sandbox Isolation},
  author={Bui, H. and Li, P. and Zafar, I. M.},
  institution={Google Research},
  year={2020}
}

@misc{owasp2025llm07,
  title={LLM07: System Prompt Leakage},
  author={{OWASP Foundation}},
  howpublished={OWASP Top 10 for LLM Applications},
  year={2025}
}

@article{alhazmi2023survey,
  title={Survey on Secure Serialization Techniques and Deserialization Vulnerabilities},
  author={Alhazmi, S. H. and Alghamdi, M. I. and Aljebali, A. and others},
  journal={IEEE Access},
  volume={11},
  year={2023}
}

@article{zhao2025protocol,
  title={Protocol Integrity Framework for AI Toolchains},
  author={Zhao, Z. and Alon, B. and Wang, K. and others},
  journal={arXiv preprint arXiv:2505.11872},
  year={2025}
}

@article{datta2025agentic,
  title={Agentic AI Security: Threats, Defenses, Evaluation, and Open Challenges},
  author={Datta, Shrestha and Nahin, Shahriar Kabir and Chhabra, Anshuman and Mohapatra, Prasant},
  journal={arXiv preprint arXiv:2510.23883},
  year={2025}
}

@article{raza2025trism,
  title={TRiSM for Agentic AI: A Review of Trust, Risk, and Security Management in LLM-based Agentic Multi-Agent Systems},
  author={Raza, Shaina and Sapkota, Ranjan and Karkee, Manoj and Emmanouilidis, Christos},
  journal={arXiv preprint arXiv:2506.04133},
  year={2025}
}

@article{delrosario2025architecting,
  title={Architecting Resilient LLM Agents: A Guide to Secure Plan-then-Execute Implementations},
  author={Del Rosario, R. F. and Krawiecka, K. and Schroeder de Witt, C.},
  journal={arXiv preprint arXiv:2509.08646},
  year={2025}
}

@article{gao2025ragguard,
  title={RAGGuard: Retrieval Augmentation with Verifiable Provenance},
  author={Gao, Y. and others},
  journal={arXiv preprint arXiv:2505.11221},
  year={2025}
}

@article{cross2025smcp,
  author = {Cross, Elena},
  title = {The "S" in {MCP} Stands for Security},
  journal = {Medium},
  year = {2025},
  month = {apr},
  url = {https://elenacross7.medium.com/%EF%B8%8F-the-s-in-mcp-stands-for-security-91407b33ed6b},
  note = {Blog Post}
}

@article{pomerium2025airoot,
  author = {{Pomerium}},
  title = {When {AI} Has Root: Lessons from the {Supabase} {MCP} Data Leak},
  journal = {Pomerium Blog},
  year = {2025},
  month = {jul},
  url = {https://www.pomerium.com/blog/when-ai-has-root-lessons-from-the-supabase-mcp-data-leak},
  note = {Post-Mortem Analysis}
}

@article{forgecode2025crisis,
  author = {{Forge Code}},
  title = {{MCP} Security Crisis: Uncovering Vulnerabilities and Attack Vectors - Part 1},
  journal = {Forge Code Blog},
  year = {2025},
  month = {jun},
  url = {https://forgecode.dev/blog/prevent-attacks-on-mcp/},
  note = {Security Guide}
}

@misc{owasp2024svs,
  author = {{OWASP Foundation}},
  title = {{OWASP} {LLM} Security Verification Standard ({SVS})},
  howpublished = {Industry Standard Draft},
  year = {2024},
  url = {https://owasp.org/www-project-llm-verification-standard/}
}

@article{tihanyi2025vulndetect,
  author = {Tihanyi, Norbert and Bisztray, Tamas and others},
  title = {Vulnerability Detection: From Formal Verification to Large Language Models},
  journal = {arXiv preprint arXiv:2503.10784},
  year = {2025},
  eprint = {2503.10784},
  archivePrefix = {arXiv},
  primaryClass = {cs.SE}
}

@article{xing2025mcpguard,
  author = {Xing, Wenpeng and Qi, Zhonghao and others},
  title = {{MCP-Guard}: A Defense Framework for Model Context Protocol Integrity in {LLM} Applications},
  journal = {arXiv preprint arXiv:2508.10991},
  year = {2025},
  eprint = {2508.10991},
  archivePrefix = {arXiv},
  primaryClass = {cs.CR}
}

@article{forgecode2025mcp2,
  author = {{Forge Code}},
  title = {{MCP} Security Prevention: Practical Strategies for {AI} Development - Part 2},
  journal = {Forge Code Blog},
  year = {2025},
  url = {https://forgecode.dev/blog/prevent-attacks-on-mcp-part2/},
  note = {Security Guide}
}

@article{huang2024longsafety,
  author = {Huang, H. and others},
  title = {{LongSafety}: Enhance Safety for Long-Context {LLMs}},
  journal = {arXiv preprint arXiv:2411.06899},
  year = {2024},
  eprint = {2411.06899},
  archivePrefix = {arXiv},
  primaryClass = {cs.LG}
}

@article{generalanalysis2024supabase,
  author = {{General Analysis}},
  title = {{Supabase} {MCP} can leak your entire {SQL} database},
  journal = {General Analysis Blog},
  year = {2024},
  url = {https://www.generalanalysis.com/blog/supabase-mcp-blog},
  note = {Incident Analysis}
}

@article{clop2024backdoored,
  author = {Clop, Cody and Teglia, Yannick},
  title = {Backdoored Retrievers for Prompt Injection Attacks on Retrieval Augmented Generation},
  journal = {arXiv preprint arXiv:2410.14479},
  year = {2024},
  eprint = {2410.14479},
  archivePrefix = {arXiv},
  primaryClass = {cs.CR}
}

@inproceedings{wu2025promptleakage,
  author = {Wu, M. and others},
  title = {I Know What You Asked: Prompt Leakage via {KV}-Cache Sharing in Multi-Tenant {LLM} Serving},
  booktitle = {Proceedings of the Network and Distributed System Security (NDSS) Symposium},
  year = {2025},
  url = {https://www.ndss-symposium.org/wp-content/uploads/2025-1772-paper.pdf}
}

@misc{owasp2024vector,
  author = {{OWASP Foundation}},
  title = {{LLM08}: Vector and Embedding Weaknesses},
  howpublished = {{OWASP} {LLM} Top 10 Risk},
  year = {2024},
  url = {https://genai.owasp.org/llmrisk/llm08-excessive-agency/}
}

@article{brimlabs2024multiuser,
  author = {{Brimlabs}},
  title = {Why {MCP} is Crucial for Building Multi-User, Multi-Tenant {LLM} Applications},
  journal = {Brimlabs Blog},
  year = {2024},
  url = {https://brimlabs.ai/blog/why-mcp-is-crucial-for-building-multi-user-multi-tenant-llm-applications/},
  note = {Architecture Guide}
}

@article{aws2024multitenant,
  author = {{Amazon Web Services (AWS)}},
  title = {Multi-tenant {RAG} implementation with {Amazon Bedrock} and {Amazon OpenSearch} Service for {SaaS} using {JWT}},
  journal = {AWS Machine Learning Blog},
  year = {2024},
  url = {https://aws.amazon.com/blogs/machine-learning/multi-tenant-rag-implementation-with-amazon-bedrock-and-amazon-opensearch-service-for-saas-using-jwt/},
  note = {Implementation Guide}
}

@article{scmedia2024vectorflaws,
  author = {{SC Media}},
  title = {{LLM} vector flaws threaten data security, privacy, and model integrity},
  journal = {SC Media},
  year = {2024},
  url = {https://www.scworld.com/feature/llm-vector-flaws-threaten-data-security-privacy-and-model-integrity},
  note = {News Article}
}

@misc{javelin2025launch,
  author = {{Javelin}},
  title = {Javelin Launches {MCP} Security to Bring Defense-in-Depth to the Model Context Protocol Layer},
  howpublished = {Press Release, BusinessWire},
  year = {2025},
  url = {https://www.businesswire.com/news/home/20250819727553/en/Javelin-Launches-MCP-Security-to-Bring-DefenseinDepth-to-the-Model-Context-Protocol-Layer},
  note = {Accessed: 2025-12-07}
}

@article{ehtesham2025survey,
  title={A survey of agent interoperability protocols: Model Context Protocol (MCP), Agent Communication Protocol (ACP), Agent-to-Agent Protocol (A2A), and Agent Network Protocol (ANP)},
  author={Ehtesham, Abul and Singh, Aditi and Gupta, Gaurav Kumar and Kumar, Saket},
  journal={arXiv preprint arXiv:2505.02279},
  year={2025}
}

\end{document}